\makeatletter
\providecommand \@ifxundefined [1]{%
 \@ifx{#1\undefined}
}%
\providecommand \@ifnum [1]{%
 \ifnum #1\expandafter \@firstoftwo
 \else \expandafter \@secondoftwo
 \fi
}%
\providecommand \@ifx [1]{%
 \ifx #1\expandafter \@firstoftwo
 \else \expandafter \@secondoftwo
 \fi
}%
\providecommand \href@noop [0]{\@secondoftwo}%
\providecommand \href [0]{\begingroup \@sanitize@url \@href}%
\providecommand \@href[1]{\@@startlink{#1}\@@href}%
\providecommand \@@href[1]{\endgroup#1\@@endlink}%
\providecommand \@sanitize@url [0]{\catcode `\\12\catcode `\$12\catcode
  `\&12\catcode `\#12\catcode `\^12\catcode `\_12\catcode `\%12\relax}%
\providecommand \@@startlink[1]{}%
\providecommand \@@endlink[0]{}%
\providecommand \url  [0]{\begingroup\@sanitize@url \@url }%
\providecommand \@url [1]{\endgroup\@href {#1}{\urlprefix }}%
\providecommand \urlprefix  [0]{URL }%
\providecommand \selectlanguage [0]{\@gobble}%
\providecommand \bibinfo  [0]{\@secondoftwo}%
\providecommand \bibfield  [0]{\@secondoftwo}%
\providecommand \BibitemShut  [1]{\csname bibitem#1\endcsname}%
\let\auto@bib@innerbib\@empty
\documentclass[aps,prc,reprint,amsmath,amssymb,nofootinbib,preprintnumbers]{revtex4-1}
\usepackage{graphicx,amsmath,amsthm,amsfonts,amssymb,xcolor}
\usepackage{multirow}
\usepackage[vcentermath]{youngtab}
\usepackage{afterpage}

\begin{document}

\preprint{BRX-TH-6290}

\title{Worldsheet Interpretation of the Level-Rank Duality}
\author{Masoud Soroush} \affiliation{Department of Mathematics, Brandeis University
\\ Waltham, MA 02453 USA\\
Martin A. Fisher School of Physics, Brandeis University\\
Waltham, MA 02453 USA}

\begin{abstract}
Level-rank duality relates the observables of two different Chern-Simons theories in which the roles of the Chern-Simons level and the rank of the gauge group are exchanged. In this note, we explore the consequences of this duality in the realm of topological string theory. We show that this duality induces a number of identities between the open Gromov-Witten invariants of the geometries associated with a knot ${\cal K}$ and its mirror image $\tilde{\cal K}$. We show how these identities arise both in the A-model and in the dual B-model. 
\end{abstract}

\maketitle

\noindent {\bf I. Introduction}
\vskip 0.2cm

The level-rank duality states that there is a correspondence between the primary fields (and also the correlation functions) of two seemingly different rational conformal field theories, namely the $SU(N)_k$ and the $SU(k)_N$ WZW models \cite{Naculich:1990hg,Naculich:1992uf}.\footnote{The level-rank duality has been promoted to other classical Lie groups as well  \cite{Naculich:1990pa,Mlawer:1990uv}.} The isomorphism between the vector space of conformal blocks of the WZW model on a surface $\Sigma$ and the Hilbert space of the quantized Chern-Simons theory on $\Sigma\times\mathbb{R}$ \cite{Witten:1988hf} lifts the level-rank duality to a duality between two different Chern-Simons theories. The consequences of this duality for the Chern-Simons observables -- which coincide with certain knot invariants \cite{Witten:1988hf} -- have been studied in a great detail in \cite{Mlawer:1990uv,Naculich:1992uf}. An important consequence of this duality is a simple relation between the colored HOMFLY invariants of a knot ${\cal K}$ and its mirror knot $\tilde{\cal K}$. 

It is well-known that Chern-Simons theory defined on a three-manifold $M$ is equivalent to the A-model open topological string theory on the total space $T^* M\xrightarrow{\pi} M$ \cite{Witten:1992fb}. The Chern-Simons partition function is written in terms of the open topological string amplitudes with the target space $T^* M\xrightarrow{\pi} M$. In case of $M=S^3$, these open-string amplitudes are efficiently captured by the closed-string amplitudes of the resolved conifold geometry via the large $N$ duality \cite{Gopakumar:1998ki}. Likewise, the Wilson loop expectation values of Chern-Simons theory are encapsulated in the A-model open topological string amplitudes on the same target space, $T^* S^3\xrightarrow{\pi} S^3$, in the presence of probe D4-branes \cite{Ooguri:1999bv}. In this manner, the large $N$ duality maps the Chern-Simons Wilson loop expectation value of a knot ${\cal K}$ to open-string amplitudes of the resolved conifold geometry with the insertion of a Lagrangian cycle associated to ${\cal K}$. \footnote{ For a review of the subject see \cite{Marino:2010wm}.} Roughly speaking, open-string amplitudes count the number of holomorphic maps from a genus $g$ Riemann surface with $h$ boundary components to the target space $O(-1)\oplus O(-1)\rightarrow {\mathbb{P}}^1$. The boundary components of the Riemann surface are mapped to the Lagrangian cycle associated to a given knot. The Lagrangian cycle provides the appropriate boundary conditions for open strings. Furthermore, mirror symmetry provides another way to approach the problem of calculating open-string amplitudes associated to the insertion of a Lagrangian cycle, by means of the dual B-model geometry. 

The large $N$ duality makes an intriguing connection between knot invariants and the open Gromov-Witten invariants in the realm of topological string theory. A natural question to ask in this context is how the level-rank duality emerges in topological string theory. The level-rank duality implies a number of identities for knot invariants. Therefore, it is natural to expect from  this duality to induce certain identities between the corresponding open Gromov-Witten invariants. In this note, we would like to address this question from both the A-model and the B-model perspectives.

From the A-model point of view, a knot ${\cal K}$ is substituted by a Lagrangian cycle ${\cal L}_{\cal K}$ in $T^* S^3\rightarrow S^3$. The construction of the Lagrangian cycle ${\cal L}_{\cal K}$ was initiated in \cite{Taubes:2001wk,Koshkin:2005rr} by constructing the conormal bundle to the knot ${\cal K}$ in the ambient target space $T^* S^3\rightarrow S^3$. However, the rigorous construction of the Lagrangian cycles associated with an algebraic knot ${\cal K}$ -- before and after the large $N$ transition --  was established in \cite{Diaconescu:2011xr}. In this construction, replacing an algebraic knot ${\cal K}$ by its mirror image $\tilde{\cal K}$ amounts to changing the Lagrangian cycle ${\cal L}_{\cal K}$ to a new Lagrangian which describes $\tilde{K}$ in the topological string setup. This would affect the relevant open string amplitudes, and hence the corresponding Gromov-Witten invariants. Tracing this change in the amplitude, we figure out the identities between the Gromov-Witten invariants associated with ${\cal K}$ and its mirror image $\tilde{\cal K}$.

In the B-model approach, the ambient space geometry is the mirror of the resolved cornfield which is a non-compact Calabi-Yau threefold. Under the dictionary of mirror symmetry, the Lagrangian cycles will be translated to holomorphic curves in the mirror ambient space. When one deals with toric Lagrangian cycles, the dictionary of the mirror symmetry is quite explicit \cite{Aganagic:2000gs,Aganagic:2001nx}, and one knows how to exactly construct the mirror curve associated with the Harvey-Lawson Lagrangian cycles. However, the Lagrangian cycles constructed in \cite{Diaconescu:2011xr} are non-toric, and the recipe of mirror symmetry is not applicable to this situation, and one cannot easily figure out the corresponding mirror curve associated with an algebraic knot ${\cal K}$. Nonetheless, one can deal with this problem in an indirect way. It has been shown that the mirror curve associated with a knot ${\cal K}$ in fact coincides with a topological invariant of the knot, known as the augmentation polynomial \cite{Aganagic:2012jb}. This quantity appears in the context of knot contact homology, and there are methods to calculate it (for a nice review of the subject see \cite{Ng:2012jx,Ekholm:2013xoa}). The higher genus and higher hole amplitudes for knots are then constructed using the techniques spelled out in a great detail in \cite{Gu:2014yba}. We trace the effect of changing a knot ${\cal K}$ with its mirror image $\tilde{\cal K}$ on the augmentation polynomial and show how the identities between open Gromov-Witten invariants of ${\cal K}$ and $\tilde{\cal K}$ are realized in the B-model. 

In this note, we proceed as follows. In section II., we show that the level-rank duality of Chern-Simons theory leads to a number of identities between the open Gromov-Witten invariants associated with a knot ${\cal K}$ and its mirror $\tilde{\cal K}$. In section III., we consider the Lagrangian cycles constructed in \cite{Diaconescu:2011xr} for an algebraic knot ${\cal K}$. We show how the Lagrangian cycles which describe ${\cal K}$ are affected when ${\cal K}$ is replaced by its mirror image $\tilde{\cal K}$. Following the change in Lagrangian cycles yields  the identities of section II between the open Gromov-Witten invariants of ${\cal K}$ and $\tilde{\cal K}$. In section IV., we illustrate how the identities found in section II are realized in the B-model approach. In appendix A., a number of open Gromov-Witten invariants with low genus and holes associated with the $(3,2)$ and $(5,3)$ torus knots and their mirrors are explicitly presented.  

\vspace{0.5cm}

\noindent {\bf II. Chern-Simons Considerations}
\vskip 0.2cm
In knot theory, there are certain topological operations which generate a new knot from a given knot. The first of these operations, $O$,  reverses the orientation of an oriented knot along it, $O({\cal K})=-{\cal K}$. The second operation, $M$, is  a reflection\footnote{ The mirror image of a knot ${\cal K}$ is independent of the choice of $M$, because the image of ${\cal K}$ under any other orientation-preserving homomorphism is ambient isotopic to $\tilde{\cal K}$ \cite{Cromwell}.} 
\begin{equation}\label{Mref}
M(x,y,z)=(x,y,-z)
\end{equation}
of the ambient space. The image of a given knot ${\cal K}$ under this map is the mirror image of ${\cal K}$, and it is denoted by $\tilde{\cal K}=M({\cal K})$. In this process, in the knot diagram of ${\cal K}$, all under-crossings are replaced by over-crossings and vice versa. The third operation, $P$, is the composition of the first two $P=O\circ M$. It turns out that $\{\mathbb{I},M,O,P\}$, in which $\mathbb{I}$ is the identity map, forms an abelian group under composition of maps. The mirror image of a given knot ${\cal K}$ is not necessarily ambient isotopic to ${\cal K}$. Although for chiral knots, a knot ${\cal K}$ and its mirror image $\tilde{\cal K}$ are topologically distinct, their knot invariants are closely related. It has been shown in \cite{Naculich:1990hg,Naculich:1992uf} that in the framework of Chern-Simons theory the relation between the invariants of a knot ${\cal K}$ and its mirror $\tilde{\cal K}$ is governed by the level-rank duality of Chern-Simons theory.   

Although exchanging the roles of the rank and the level of an affine Lie algebra $G(N)_k$ is not a symmetry of the algebra, it was shown that $G(N)_k$ and $G(k)_N$ Chern-Simons theories are dual to each other \cite{Naculich:1990hg,Naculich:1990pa}. Specifically for $SU(N)_k$ and $SU(k)_N$ Chern-Simons theories defined on $S^3$, the observables of the two theories are in a one-to-one correspondence 
\begin{equation}\label{Wilson}
\langle W^{(\tilde{\cal K})}_{\mu}\rangle_{SU(N)_{k}}=\langle W^{({\cal K})}_{\mu^t}\rangle_{SU(k)_{N}}
\end{equation}
In (\ref{Wilson}), $\langle W^{({\cal K})}_{\mu}\rangle_{SU(N)_{k}}$ is the expectation value of the Wilson operator around a  knot ${\cal K}$, and the Young tableau $\mu$ specifies an irreducible highest weight representation of $SU(N)_{k}$. The Young tableau $\mu^{t}$ is the transpose of $\mu$, and $\tilde{\cal K}$ is the mirror of ${\cal K}$. In order to make contact with knot invariants, it is convenient to work with the two following parameters 
\begin{equation}
q=\exp\Big(\frac{2\pi i}{k+N}\Big)\quad,\quad Q=\exp\Big(\frac{2\pi iN}{k+N}\Big) 
\end{equation}
The HOMFLY invariant associated to a knot ${\cal K}$, colored with representation $\mu$, is related to the Wilson loop expectation value along that knot in representation $\mu$ 
\begin{equation}
{\cal H}^{({\cal K})}_{\mu}(Q,q)=\frac{\langle W^{({\cal K})}_{\mu}\rangle_{U(N)_k}}{S_{00}}
\end{equation}
where $S_{00}$ is the partition function of Chern-Simons theory on $S^3$. Under the exchange $k\leftrightarrow N$, the quantum parameter $q$ remains invariant $q\leftrightarrow q$, and $Q$ is replaced by its inverse $Q\leftrightarrow Q^{-1}$. After incorporating the contribution of the $U(1)$ factor, the level-rank duality (\ref{Wilson}) implies
\begin{equation}\label{LRdual}
{\cal H}^{(\tilde{\cal K})}_{\mu}(Q,q)=(-1)^{|\mu|}\,{\cal H}^{({\cal K})}_{\mu^{t}}(Q^{-1},q)
\end{equation}
We would like to explore the consequences of (\ref{LRdual}) for open topological string amplitudes. It is well known that Chern-Simons theory on $S^3$ is equivalent to topological A-model with the target $T^* S^3\rightarrow S^3$ with Lagrangian boundary conditions \cite{Witten:1992fb}. Roughly speaking, open string amplitudes count the number of holomorphic maps from the worldsheet -- which is a genus $g$ Riemann surface with $h$ boundary components -- to the target space $T^* S^3\rightarrow S^3$. In this correspondence $q=e^{ig_s}$ and $Q=e^{t}$, where $t=iNg_s$ is the 't Hooft coupling constant. The genus $g$ amplitude with $h$ boundaries and winding vector $\vec{k}$ associated to a knot ${\cal K}$ is given in terms of the HOMFLY invariants of ${\cal K}$ in the following way 
\begin{equation*}
\omega_{(g,h,\vec{k})}^{({\cal K})}(Q)=\Bigg[\Big[\log \big(\sum_{\vec{k}'}Z_{\vec{k}'}^{({\cal K})}(Q,q)\, \mbox{Tr}_{\vec{k}'}V\big)\Big]_{\mbox{Tr}_{\vec{k}}V}\Bigg]_{g_s^{2g+h-2}}
\end{equation*}
In the above expression, after the genus expansion, we only pick the coefficient of $g_s^{2g+h-2}$. The partition function $Z_{\vec{k}}^{({\cal K})}$ is given by
\begin{equation*}
Z_{\vec{k}}^{({\cal K})}(Q,q)=\sum_{|\mu|=\ell(\vec{k})}\chi_{\mu}(C_{\vec{k}})\,{\cal H}^{({\cal K})}_{\mu}(Q,q)
\end{equation*}
in which $\chi_{\mu}(C_{\vec{k}})$ is the character of the symmetric group ${\cal S}_{\ell(\vec{k})}$ with conjugacy class $C_{\vec{k}}$. Since the colored HOMFLY invariants of ${\cal K}$ are all polynomials (after multiplication by an appropriate factor $Q^{\ell}$, where $\ell$ is either an integer or a half-integer number) in terms of $Q$, the amplitude $\omega_{(g,h,\vec{k})}^{({\cal K})}(Q)$ is also a polynomial of $Q$. The open Gromov-Witten invariants of ${\cal K}$ and the mirror knot $\tilde{\cal K}$ are then obtained as
\begin{equation}\label{Amps}
\begin{aligned}
&\omega_{(g,h,\vec{k})}^{({\cal K})}(Q)=\sum_{d=d_{\tiny\mbox{{min}}}}^{d_{\tiny{\mbox{max}}}}GW_{(g,h)}^{({\cal K})}(d,\vec{k})\,Q^{d}\ ,\\
&\omega_{(g,h,\vec{k})}^{(\tilde{{\cal K}})}(Q)=\sum_{d=d_{\tiny\mbox{{min}}}}^{d_{\tiny{\mbox{max}}}}GW_{(g,h)}^{(\tilde{{\cal K}})}(d,\vec{k})\,Q^{d}\ .
\end{aligned}
\end{equation}
where $d_{\tiny\mbox{{min}}}$ and $d_{\tiny\mbox{{max}}}$ are the lowest and highest degrees of non-vanishing Gromov-Witten invariants associated with the genus $g$ amplitude with $h$ boundaries and winding $\vec{k}$ respectively. Because of (\ref{LRdual}), it is evident that there should be relations between the open Gromov-Witten invariants associated with ${\cal K}$ and $\tilde{\cal K}$. To figure out this relationship, let us first find how the two partition functions are related
\begin{equation}\label{Pfunc}
\begin{aligned}
Z^{(\tilde{\cal K})}_{\vec{k}}&(Q,q)=\sum_{|\mu|=\ell(\vec{k})}\chi_{\mu}(C_{\vec{k}})\,{\cal H}^{(\tilde{\cal K})}_{\mu}(Q,q)\\
&=(-1)^{|\vec{k}|}\sum_{|\mu|=\ell(\vec{k})}\chi_{\mu^t}(C_{\vec{k}}){\cal H}^{({\cal K})}_{\mu^t}(Q^{-1},q)\\
&=(-1)^{|\vec{k}|}Z^{({\cal K})}_{\vec{k}}(Q^{-1},q)
\end{aligned}
\end{equation}
In the second line of (\ref{Pfunc}), we have used (\ref{LRdual}), and the fact that
\begin{equation}
\chi_{\mu^t}(C_{\vec{k}})=(-1)^{|\vec{k}|+\ell(\vec{k})}\,\chi_{\mu}(C_{\vec{k}})
\end{equation}
where $|\vec{k}|$ and $\ell(\vec{k})$ are the number of holes and the total winding associated with the winding vector $\vec{k}$ respectively. To find the relation between the amplitudes associated with ${\cal K}$ and $\tilde{\cal K}$, we notice that $\mbox{Tr}_{\vec{k}_{1}}V\cdot\mbox{Tr}_{\vec{k}_{2}}V=\mbox{Tr}_{\vec{k}_{1}+\vec{k}_2}V$, and the fact that $|\vec{k}|_{1}+|\vec{k}_{2}|=|\vec{k}_{1}+\vec{k}_{2}|$. These considerations together with (\ref{Amps}) and (\ref{Pfunc}) result in the following simple identity between the open Gromov-Witten invariants associated with a knot ${\cal K}$ and its mirror $\tilde{\cal K}$ \footnote{ For the case of unknot, the special case of this relation for disks ($g=0$, $h=1$) was proved in \cite{Marino:2001re}, using techniques of mirror symmetry. Of course, the mirror image of unknot is the unknot itself, and this relation becomes a relation between disk invariants of unknot with different degrees.}
\begin{equation}\label{GWrel}
\boxed{GW_{(g,h)}^{(\tilde{\cal K})}(d,\vec{k})=(-1)^{h}GW_{(g,h)}^{({\cal K})}(d_{\tiny\mbox{{min}}}+d_{\tiny\mbox{{max}}}-d,\vec{k})}   
\end{equation}
The factor $(-1)^h$ has a physical interpretation in terms of topological amplitudes. Once a knot ${\cal K}$ is exchanged by its mirror $\tilde{\cal K}$, the orientation of the Lagrangian cycle associated with ${\cal K}$ is reversed. This in particular implies that the orientation of all $h$ circles of the corresponding amplitude, which end on the Lagrangian cycle, are reversed too. In string theory language, this parity operation on the Lagrangian brane changes the action by an overall minus sign. Therefore a genus $g$ topological amplitude with $h$ boundary components on the corresponding D-brane is modified by a factor of $(-1)^h$.

Explicit examples of Gromov-Witten invariants associated with the (3,2) and (5,3) torus knots and their mirrors have been presented in appendix A.

\vspace{0.5cm}
\noindent {\bf III. A-model}
\vskip 0.2cm

In this section, we would like to see how identity (\ref{GWrel}) arises in the topological A-model. In the A-model, a knot ${\cal K}$ is substituted by a Lagrangian cycle. The recipe for constructing the correct Lagrangian cycle associated to a general knot ${\cal K}$ is not yet known. However, for a large class of knots, known as algebraic knots, the Lagrangian cycles were delicately constructed in \cite{Diaconescu:2011xr}, based on the previous works \cite{Taubes:2001wk,Koshkin:2005rr,Labastida:2000yw}. Although the Lagrangian cycles associated with algebraic knots are known, due to their complicated nature, one can explicitly compute topological amplitudes (using localization techniques of \cite{Kontsevich:1994na,Graber:2001dw,Katz:2001vm,Li:2001sg}) only for a subclass of algebraic knots which preserve certain ${\mathbb{C}}^*$ symmetries. This subclass consists of torus knots. In this section, we first explain what happens to the Lagrangian cycles associated to an algebraic knot if one substitutes a knot ${\cal K}$ by its mirror $\tilde{\cal K}$. In the second step, we show how this change leads to identity (\ref{GWrel}) for torus knots.  

An algebraic knot is defined in the ${\mathbb{C}}^2$-plane. Let $x$ and $y$ locally describe the ${\mathbb{C}}^2$-plane in consideration. An algebraic knot is then defined as the intersection loci of the holomorphic curve 
\begin{equation}\label{AlgK}
F(x,y)=0
\end{equation}
with the three-sphere $S^{3}=\{(x,y)\in{\mathbb{C}}^2|\, |x|^{2}+|y|^{2}=2\}$. For instance an $(r,s)$ torus knot -- in which $r$ and $s$ are coprime numbers -- is specified by
\begin{equation}
F^{(r,s)}(x,y)=x^r-y^s
\end{equation}
To see why the intersection of the holomorphic curve $F^{(r,s)}(x,y)=0$ with the three-sphere $S^3$ is a $(r,s)$ torus knot, let us define the following Clifford torus $T^2=S^{1}_{\theta}\times S^{1}_{\phi}$ inside the above three-sphere, in which
\begin{equation}
\begin{aligned}
&S^{1}_{\theta}=\{(x,y)|\,x=e^{i\theta},y=0,\,0\leq\theta <2\pi\}\\
&S^{1}_{\phi}=\{(x,y)|\,x=0,y=e^{i\phi},\,0\leq\phi<2\pi\}
\end{aligned}
\end{equation}
When $r\theta-s\phi=2\pi k$ for any $k\in {\mathbb{Z}}$, we find a torus knot on the surface of the Clifford torus. Without loss of generality we set $k=0$, and we define $s\varphi\equiv\theta$. Then a $(r,s)$ torus knot is parametrized as
\begin{equation*}
{\cal K}^{(r,s)}=\{(x,y)\in{\mathbb{C}}^2|x=e^{is\varphi}, y=e^{ir\varphi}, 0\leq\varphi<2\pi\}
\end{equation*}
The mirror image of an algebraic knot ${\cal K}$, defined by (\ref{AlgK}), is obtained by the operation (\ref{Mref}). In this setup, the mirror reflection is equivalent to substituting one of the coordinates of the ambient ${\mathbb{C}}^2$-plane by its complex conjugate. Therefore, the mirror knot $\tilde{\cal K}$ is defined by
\begin{equation}
F(x,\bar{y})=0
\end{equation}
where $\bar{y}$ is the complex conjugate of $y$. This implies that the mirror of a $(r,s)$ torus knot is parametrized as
\begin{equation*}
\tilde{\cal K}^{(r,s)}=\{(x,y)\in{\mathbb{C}}^2|x=e^{is\varphi}, y=e^{-ir\varphi}, 0\leq\varphi<2\pi\}
\end{equation*}
Comparing $\tilde{\cal K}^{(r,s)}$ with ${\cal K}^{(r,s)}$, it is clear that the mirror of a $(r,s)$ torus knot is equivalent to the $(-r,s)$ torus knot; confirming a well known fact in knot theory. 

Now, we would like to construct the Lagrangian cycles associated with an algebraic knot and its mirror, following \cite{Diaconescu:2011xr}. The deformed conifold space $X_{\mu}$ is defined as a hypersurface in ${\mathbb{C}}^4$ 
\begin{equation}
X_{\mu}=\{(x,y,z,w)\in{\mathbb{C}}^4 |\, xz-yw=\mu\}
\end{equation}
which is equipped with the natural symplectic two-form induced from the ambient ${\mathbb{C}}^4$. $X_{\mu}$ is isomorphic to the total space $T^* S^3\xrightarrow{\pi} S^3$. Defining $z_1=(x+z)/2$, $z_2=-i/2(x-z)$, $z_3=-1/2(y-w)$, and $z_4=i/2(y+w)$, we realize $\sum_i z_i^2=\mu$. Splitting $z_i$ into real and imaginary parts $z_i=x_i+iy_i$ ($x_i,y_i\in{\mathbb{R}}$), it is easy to see that
\begin{equation}
\vec{x}\cdot\vec{y}=0\quad,\quad |\vec{x}|^2+|\vec{y}|^2=\mu
\end{equation}
Defining the symplectomorphism $\phi_{\mu}:X_{\mu}\rightarrow X$
\begin{equation}
\phi_{\mu}(\vec{x},\vec{y})\equiv(\vec{u},\vec{v})=\Big(\frac{\vec{x}}{|\vec{x}|},-|\vec{x}|\,\vec{y}\Big)
\end{equation}
the total space $T^* S^3\rightarrow S^3$ is realized as a subspace of ${\mathbb{R}}^4 \times{\mathbb{R}}^4$
\begin{equation}
|\vec{u}|=1\quad,\quad \vec{u}\cdot\vec{v}=0
\end{equation}
in which the first equation specifies the three-sphere base, and $\vec{v}$ is the normal direction. 

To construct the Lagrangian associated with an algebraic knot ${\cal K}$, it is crucial for the Lagrangian ${\cal L}_{\cal K}$ to avoid the singular point of the conifold as it goes through the large $N$ transition \cite{Diaconescu:2011xr}. To achieve this, one has to consider a lift of the knot ${\cal K}$. A lift is an embedding of a circle $\gamma$ in $X$ (${\gamma}_{\cal K}: S^1\rightarrow X$) such that the composition $\pi\circ{\gamma}_{\cal K}: S^1\rightarrow S^3$ is the knot ${\cal K}$. In order for ${\cal K}$ to avoid self-intersection, it is also important that ${\gamma}_{\cal K}$ intersects each fiber only once. Assuming a lift $\gamma_{\cal K}$ exists
\begin{equation}
(\vec{u},\vec{v})=(\vec{f}(\theta),\vec{g}(\theta))\quad,\quad \theta\in[0,2\pi]
\end{equation}
then the Lagrangian cycle $L_{\gamma}$ is constructed by
\begin{equation}
\vec{u}=\vec{f}(\theta)\quad,\quad \dot{\vec{f}}(\theta)\cdot\big(\vec{v}-g(\theta)\big)=0
\end{equation}
The Lagrangian cycle associated to ${\cal K}$ in the deformed conifold $X_{\mu}$ is then given by ${\cal L}_{\cal K}=\phi_{\mu}^{-1}(L_{{\gamma}_{\cal K}})$. An explicit lift for algebraic knots constructed in \cite{Diaconescu:2011xr}. The starting point for $\gamma_{\cal K}$ is the holomorphic curve $Z_{\mu}^{({\cal K})}\subset X_{\mu}$
\begin{equation}
Z_{\mu}^{({\cal K})}=\{F(x,y)=0\}\cap\{F(z,-w)=0\}
\end{equation}
We notice that $Z_{\mu}^{({\cal K})}\cap S_{\mu}={\cal K}$ in which $S_{\mu}$ is the vanishing cycle of the deformed conifold. $Z_{\mu}^{({\cal K})}$ may have several distinct connected components. We take one connected component $C_{\mu}^{({\cal K})}\subset Z_{\mu}^{({\cal K})}$ which has nontrivial intersection with $S_{\mu}$. The last ingredient in construction of the lift $\gamma_{\cal K}$ is the $S^2$-bundle $P_a=\{(\vec{u},\vec{v})\in X|\,|\vec{v}|=a\}$ where $a\in {\mathbb{R}}_{+}$ \cite{Diaconescu:2011xr}. The lift $\gamma_{\cal K}$ is then given as the intersection of the sphere bundle $P_a$ and the image of $C_{\mu}^{({\cal K})}$ under the symplectomorphism $\phi_{\mu}$
\begin{equation}
\gamma_{\cal K}=P_a\cap \phi_{\mu}(C_{\mu}^{({\cal K})})
\end{equation}
As is clear from the construction of the Lagrangian cycle ${\cal L}_{\cal K}$, the choice of the knot ${\cal K}$ enters through the lift $\gamma_{\cal K}$. Therefore, in the construction of the Lagrangian cycle associated with the mirror image of ${\cal K}$, we need to adjust the corresponding lift accordingly. The only difference in the construction of the lift $\gamma_{\tilde{\cal K}}$ enters through the curve $Z_{\mu}$. For the mirror knot $\tilde{\cal K}$, the corresponding curve will be 
\begin{equation}
Z_{\mu}^{(\tilde{\cal K})}=\{F(x,\bar{y})=0\}\cap\{F(z,-\bar{w})=0\}
\end{equation}
Choosing one connected component of $Z_{\mu}^{(\tilde{\cal K})}$, the lift associated with $\tilde{\cal K}$ is given by
\begin{equation}
\gamma_{\tilde{\cal K}}=P_a\cap \phi_{\mu}(C_{\mu}^{(\tilde{\cal K})})
\end{equation}

As of now, the explicit calculation of topological string amplitudes associated with the above Lagrangian cycles is only possible for torus knots which preserve certain ${\mathbb{C}}^*$ action. In here, in the remaining part of this section, we show that the open Gromov-Witten invariants of ${\cal L}_{{\cal K}^{(r,s)}}$ and ${\cal L}_{\tilde{\cal K}^{(r,s)}}$ are related, when ${\cal K}^{(r,s)}$ is a $(r,s)$ torus knot. We notice that ${\cal L}_{{\cal K}^{(r,s)}}$ and ${\cal L}_{\tilde{\cal K}^{(r,s)}}$ preserve different ${\mathbb{C}}^*$ symmetries
\begin{equation}
\begin{aligned}
&(x,y,z,w)\mapsto (e^{is\varphi}x,e^{ir\varphi}y,e^{-is\varphi}z,e^{-ir\varphi}w)\\
&(x,y,z,w)\mapsto (e^{is\varphi}x,e^{-ir\varphi}y,e^{-is\varphi}z,e^{ir\varphi}w)
\end{aligned}
\end{equation} 
where the first and the second transformations are the symmetries of ${\cal L}_{{\cal K}^{(r,s)}}$ and ${\cal L}_{\tilde{\cal K}^{(r,s)}}$ respectively. In order to carry out the localization computation on the corresponding moduli space of stable maps with boundary components, we have to trace the above ${\mathbb{C}}^*$ symmetries in the resolved conifold. It is easy to see that the corresponding ${\mathbb{C}}^*$ symmetries that the proper Lagrangian cycles (${\cal L}_{\cal K}$ and ${\cal L}_{\tilde{\cal K}}$ after large $N$ transition transform into new Lagrangian cycles whose construction has been spelled out in detail in \cite{Diaconescu:2011xr}) preserve take the following form 
\begin{equation}\label{}
\begin{aligned}
&(x,y,z,w;[u_1,u_2])\mapsto\\
&(e^{is\varphi}x,e^{\pm ir\varphi}y,e^{-is\varphi}z,e^{\mp ir\varphi}w;[e^{-i(s\pm r)\varphi}u_1,u_2])
\end{aligned}
\end{equation}
where $[u_1,u_2]$ are the homogeneous coordinates of the ${\mathbb{P}}^1$ cycle after the small resolution. In above equation, the first and the second choices of signs correspond to the circle actions associated with ${\cal K}^{(r,s)}$ and $\tilde{\cal K}^{(r,s)}$ respectively. This choice of signs would also affect the weights of the equivariant classes of the localization computation. There are two ways to proceed. We can carry out the localization computation for amplitudes case by case for both Lagrangian cycles associated with ${\cal K}^{(r,s)}$ and $\tilde{\cal K}^{(r,s)}$ with respect to their ${\mathbb{C}}^*$ symmetries, and verify (\ref{GWrel}). We have checked (\ref{GWrel}) for several amplitudes with low genus and winding, and the results are in agreement with (\ref{GWrel}). However, we can draw a more general conclusion. It has been shown in \cite{Diaconescu:2011xr}, by relating the Gromov-Witten invariants of torus knots to those of unknot, that one can reproduce the Rosso-Jones formula \cite{RJ} for the HOMFLY polynomial of torus knots. We do knot repeat this derivation in here. Following the same line of argument as in \cite{Diaconescu:2011xr}, we can show that the generating functions for open Gromov-Witten invariants of a $(r,s)$ torus knot and its mirror image for winding one are given by
\begin{equation}
\begin{aligned}
F^{(r,s)}_1(Q,g_s)=&(-1)^{s-1}Q^{-s/2}\\
&\sum_{|\mu|=s}\chi_{\mu}(C_{k^{(s)}})e^{i\frac{r}{2s}\kappa_{\mu}g_s}\mbox{dim}_{q}\mu\\
F^{\widetilde{(r,s)}}_1(Q,g_s)=&(-1)^{s-1}Q^{-s/2}\\
&\sum_{|\mu|=s}\chi_{\mu}(C_{k^{(s)}})e^{-i\frac{r}{2s}\kappa_{\mu}g_s}\mbox{dim}_{q}\mu
\end{aligned}
\end{equation} 
These equation are precisely special cases of the Rosso-Jones formula for a $(r,s)$ torus knot and its mirror image $\widetilde{(r,s)}$. On the other hand, we know that the HOMFLY invariants of a knot and its mirror are related by (\ref{LRdual}). As a result, the identity (\ref{GWrel}) between Gromov-Witten invariants of a knot ${\cal K}$ and its mirror image $\tilde{\cal K}$ is realized from the A-model point of view for torus knots. Although $F^{(r,s)}_1$ and $F^{\widetilde{(r,s)}}_1$ are only the the one-point functions (with winding one for the boundary) associated to a $(r,s)$ torus knot and its mirror $\widetilde{(r,s)}$, this derivation is generalized to higher hole and higher winding Gromov-Witten invariants in a straightforward manner. 

Before closing this section, we would like to comment on the derivation of \cite{Jockers:2012pz} for the HOMFLY invariants of torus knots in the framework of A-model. In \cite{Jockers:2012pz}, it was shown how one can calculate the HOMFLY invariants of torus knots using the formalism of the topological vertex \cite{Aganagic:2003db}. Topological vertex calculates open topological string amplitudes associated with toric Lagrangian cycles embedded in a toric Calabi-Yau threefold. Taking the advantage of the fact that a torus knot is generated by an $SL(2,{\mathbb{Z}})$ transformation of the unknot, it was shown that the HOMFLY invariants of torus knots can be captured by performing an appropriate $SL(2,{\mathbb{Z}})$ transformation on the toric Lagrangian cycle associated with the unknot. In this derivation the framing of the toric Lagrangian cycle is a fraction of the framing of the torus knot. The standard framings for ${\cal K}^{(r,s)}$ and $\tilde{\cal K}^{(r,s)}$ torus knots are $rs$ and $-rs$ respectively. It was then shown that this sign difference leads exactly to equation (\ref{LRdual}), specialized to a $(r,s)$ torus knot and its mirror image.  

\vspace{0.5cm}
\noindent {\bf IV. B-model}
\vskip 0.2cm

In the B-model approach, the open Gromov-Witten invariants associated with toric geometries are calculated as follows. One first calculates the disk instanton numbers from an algebraic curve, known as the mirror curve. The mirror curve is a non-compact Riemann surface, and is the generating function of disk instantons. In the second step, the annulus instanton numbers are obtained from the Bergman kernel of the mirror curve, which is determined by the topology and geometry of the curve. Disk and annulus instanton numbers are the ingredients for higher invariants. The higher genus and higher hole invariants are determined by the recursive procedure of \cite{Bouchard:2007ys}.

In the present case, the situation is more subtle. First of all, since the Lagrangian cycles in the A-model are non-toric, we cannot directly obtain the corresponding mirror curves which determine the disk instantons. However, it was shown in \cite{Aganagic:2012jb} that the mirror curve associated with a knot ${\cal K}$ coincides with one of the topological invariants associated with ${\cal K}$, namely its augmentation polynomial.\footnote{ In general, the augmentation polynomial may be one irreducible component of the mirror curve.} The augmentation polynomial\footnote{ For an introduction to the subject see \cite{Ng:2012jx,Ekholm:2013xoa}.} is an algebraic curve and it reproduces the A-polynomial of ${\cal K}$ in a certain limit. It turns out that the augmentation polynomial can be constructed only by the knowledge of the HOMFLY polynomials of ${\cal K}$, colored in totally symmetric representations. Suppose $x$ and $y$ are the local coordinates on the mirror curve ${\cal A}$ (augmentation polynomial). The curve can then be represented in the following way \cite{Aganagic:2012jb,Jockers:2012pz} 
\begin{equation}\label{A-curve}
y(x)=\exp \Big(x\frac{d}{dx}\lim_{g_s\rightarrow0}\log\sum_{k=0}^{\infty}{\cal H}^{({\cal K})}_{S_k}(Q,q)\,x^k\Big)
\end{equation}
where $S_{k}$ is the $k$-th totally symmetric representation displayed by a Young diagram with total number of $k$ boxes sitting in a single row, and $q=e^{ig_s}$. In the topological string setup, the parameter $Q$ corresponds to the area of the base ${\mathbb{P}}^1$ of the resolved conifold geometry. For instance, the augmentation polynomial associated with the trefoil knot turns out to be
\begin{equation}\label{Tre-aug}
\begin{aligned}
&(1-Qy)+(y^3-y^4+2y^5-2Qy^5\\
&-Qy^6+Q^2y^7)x-y^9(1-y)x^2=0
\end{aligned}
\end{equation}
Before, discussing higher invariants in the B-model, let us see how (\ref{GWrel}) is realized in the B-model at the level of disks. In order to answer this question, we first notice that the disk instanton numbers associated with ${\cal K}$ can also be constructed by only the knowledge of HOMFLY invariants of ${\cal K}$ in totally anti-symmetric representations 
\begin{equation}\label{ASymm}
y(x)=\exp \Big(-x\frac{d}{dx}\lim_{g_s\rightarrow0}\log\sum_{k=0}^{\infty}{\cal H}^{({\cal K})}_{\Lambda_k}(Q,q)\,x^k\Big)
\end{equation}
In (\ref{ASymm}), $\Lambda_{k}$ is the $k$-th totally anti-symmetric representation displayed by a Young diagram with total number of $k$ boxes sitting in a single column. Now, let $\tilde{x}$ and $\tilde{y}$ be the local coordinates on the mirror curve $\tilde{\cal A}$ associated with the mirror knot $\tilde{\cal K}$. Similar to ${\cal K}$, the mirror curve associated to $\tilde{\cal K}$ can be constructed via (\ref{A-curve}) as
\begin{equation}\label{At-curve}
\tilde{y}(\tilde{x})=\exp \Big(\tilde{x}\frac{d}{d\tilde{x}}\lim_{g_s\rightarrow0}\log\sum_{k=0}^{\infty}{\cal H}^{(\tilde{\cal K})}_{S_k}(Q,q)\,\tilde{x}^k\Big)
\end{equation}
Using (\ref{LRdual}), we can rewrite (\ref{At-curve}) in terms of the HOMFLY invariants of the original knot ${\cal K}$ in totally anti-symmetric representations
\begin{equation*}
\tilde{y}(\tilde{x})=\exp \Big(-\tilde{x}\frac{d}{d\tilde{x}}\lim_{g_s\rightarrow0}\log\sum_{k=0}^{\infty}{\cal H}^{(\cal K)}_{\Lambda_k}(Q^{-1},q)\,(-\tilde{x})^k\Big)
\end{equation*}
Comparing the above formula with (\ref{ASymm}), and noticing the fact that the HOMFLY invariants ${\cal H}_{\mu}^{({\cal K})}(Q,q)$ are polynomials in terms of $Q$, we arrive at
\begin{equation}
GW_{(0,1)}^{(\tilde{\cal K})}(d,k)=-GW_{(0,1)}^{({\cal K})}(d_{\tiny\mbox{{min}}}+d_{\tiny\mbox{{max}}}-d,k)
\end{equation}
in which $d_{\tiny\mbox{{min}}}$ and $d_{\tiny\mbox{{max}}}$ are the lowest and the highest degrees of non-vanishing disk Gromov-Witten invariants for a given winding $k$ respectively. It is also evident that the augmentation polynomial $\tilde{\cal A}$ associated with $\tilde{\cal K}$ is simply obtained from ${\cal A}$ by sending $x\rightarrow -x$ and $Q\rightarrow Q^{-1}$. For instance for the mirror of the trefoil knot, we obtain the corresponding augmentation polynomial from (\ref{Tre-aug}) to be
\begin{equation}\label{TreM-aug}
\begin{aligned}
&Q(Q-\tilde{y})-(Q^2\tilde{y}^3-Q^2\tilde{y}^4+2Q^2\tilde{y}^5-2Q\tilde{y}^5\\
&-Q\tilde{y}^6+\tilde{y}^7)\tilde{x}-Q^2\tilde{y}^9(1-\tilde{y})\tilde{x}^2=0
\end{aligned}
\end{equation} 

Now, we turn into higher invariants. The next step is to determine the annulus instanton numbers associated to ${\cal K}$. Unlike the toric cases, the annulus numbers in this case are not given by the Bergman kernel associated with the augmentation polynomial \cite{Gu:2014yba}. However, it was shown in \cite{Gu:2014yba} that there exists another symmetric bi-differential which is the generating function of all annulus instanton numbers. It was shown in \cite{Gu:2014yba} how one can explicitly construct this bi-differential (known as the physical annulus kernel) for torus knots. This annulus kernel together with the augmentation polynomial are the ingredients for a recursive procedure to determine all other higher invariants. To see how the annulus instanton numbers of ${\cal K}$ and $\tilde{\cal K}$ are related, we first notice that all annuli can be extracted by only the knowledge of HOMFLY invariants colored with at most rows \cite{Gu:2014yba}. For brevity, let us concentrate on a specific annulus with winding vector $\vec{k}=\{1,1\}$. One can obtain the instanton numbers associated with this annulus in the following way
\begin{equation*}
\begin{aligned}
\omega&_{(0,2,\{1,1\})}^{({\cal K})}(Q)=2\Big[{\cal H}^{({\cal K})}_{\yng(3)}(Q,q)+{\cal H}^{({\cal K})}_{\yng(2,1)}(Q,q)-{\cal H}^{({\cal K})}_{\yng(1)}(Q,q)\\
&(2{\cal H}^{({\cal K})}_{\yng(2)}(Q,q)+{\cal H}^{({\cal K})}_{\yng(1,1)}(Q,q)-{\cal H}^{({\cal K})}_{\yng(1)}(Q,q)^2)\Big]_{{\cal O}(g_s^0)}
\end{aligned}
\end{equation*}
Equivalently, it turns out that we can obtain all annulus instanton numbers only with the knowledge of HOMFLY invariants with at most two columns. For the annulus $\vec{k}=\{1,1\}$, we have
\begin{equation*}
\begin{aligned}
&\omega_{(0,2,\{1,1\})}^{({\cal K})}(Q)=-2\Big[{\cal H}^{({\cal K})}_{\yng(1,1,1)}(Q,q)+{\cal H}^{({\cal K})}_{\yng(2,1)}(Q,q)-\\
&{\cal H}^{({\cal K})}_{\yng(1)}(Q,q)({\cal H}^{({\cal K})}_{\yng(2)}(Q,q)+2{\cal H}^{({\cal K})}_{\yng(1,1)}(Q,q)-{\cal H}^{({\cal K})}_{\yng(1)}(Q,q)^2)\Big]_{{\cal O}(g_s^0)}
\end{aligned}
\end{equation*}
The same expressions would hold for the annulus numbers of the mirror knot $\tilde{\cal K}$ if one uses the HOMFLY invariants of $\tilde{\cal K}$ accordingly. If we use (\ref{LRdual}), we easily realize that
\begin{equation*}
\begin{aligned}
\omega&_{(0,2,\{1,1\})}^{(\tilde{\cal K})}(Q)=2\Big[{\cal H}^{({\cal K})}_{\yng(3)}(Q^{-1},q)+{\cal H}^{({\cal K})}_{\yng(2,1)}(Q^{-1},q)\\
&-{\cal H}^{({\cal K})}_{\yng(1)}(Q^{-1},q)(2{\cal H}^{({\cal K})}_{\yng(2)}(Q^{-1},q)+{\cal H}^{({\cal K})}_{\yng(1,1)}(Q^{-1},q)\\
&-{\cal H}^{({\cal K})}_{\yng(1)}(Q^{-1},q)^2)\Big]_{{\cal O}(g_s^0)}
\end{aligned}
\end{equation*}
Noticing that the HOMFLY invariants are polynomials in terms of $Q$, we arrive at
\begin{equation*}
GW_{(0,2)}^{(\tilde{\cal K})}(d,\{1,1\})=GW_{(0,2)}^{({\cal K})}(d_{\tiny\mbox{{min}}}+d_{\tiny\mbox{{max}}}-d,\{1,1\})
\end{equation*}
where $d_{\tiny\mbox{{min}}}$ and $d_{\tiny\mbox{{max}}}$ are the lowest and the highest degrees of non-vanishing annulus Gromov-Witten invariants for the  winding $\vec{k}=\{1,1\}$ respectively. This argument is easily generalized to any windings for the two boundary components of an annulus instanton, and we find 
\begin{equation*}
GW_{(0,2)}^{(\tilde{\cal K})}(d,\vec{k})=GW_{(0,2)}^{({\cal K})}(d_{\tiny\mbox{{min}}}+d_{\tiny\mbox{{max}}}-d,\vec{k})
\end{equation*}
in which $\vec{k}=\{0,0,\cdots,1,0,0,\cdots,1\}$ is the winding vector associated to a general annulus amplitude. 

The higher genus and higher hole invariants associated with ${\cal K}$ and $\tilde{\cal K}$ are obtained through the recursive procedure of \cite{Eynard:2007kz,Bouchard:2007ys}, with the kernels constructed in \cite{Gu:2014yba}.
\begin{equation*}
\begin{aligned}
&\Omega_{(g,h+1)}^{({\cal K})}(p,J)=\sum_{i}\mathop{\mbox{Res}}_{q\rightarrow a_i}K(p,q)\Big(\Omega_{(g-1,h+2)}^{({\cal K})}(q,\bar{q},J)\\
&+\sum_{\ell=0}^g\sum_{I\subset J}^{'}\Omega^{({\cal K})}_{(\ell,| I |+1)}(q,I)\Omega_{(g-\ell,h-| I |+1)}^{({\cal K})}(\bar{q},J\backslash I)
\end{aligned}
\end{equation*}
In above, the differentials $\Omega^{({\cal K})}_{(g,h)}$ are the generating functions of genus $g$ amplitudes with $h$ boundary components\footnote{ For more details on notations, consult \cite{Gu:2014yba}.} 
\begin{equation}
\begin{aligned}
\Omega^{({\cal K})}_{(g,h)}&(x_{1},\cdots,x_{h})=\sum_{|\vec{k}|=h}\omega_{(g,h,\vec{k})}(Q)\\
&\Big(x_{1}^{k_1}\cdots x_{h}^{k_h}+\mbox{permutations}\Big)dx_1\cdots dx_h
\end{aligned}
\end{equation}
The recursion kernel $K(p,q)$ is made of the physical annulus kernel and the canonical meromorphic one-form which generates disk instantons. It is easily seen that by sending ${\cal K}\rightarrow\tilde{\cal K}$, the recursion kernel gets a minus sign and its $Q$ dependence is substituted by $Q^{-1}$ in it. Then, by induction, one recognizes that the identity (\ref{GWrel}) is fulfilled among the Gromov-Witten invariants of ${\cal K}$ and $\tilde{\cal K}$.  

Before we conclude this section, there are three comments in order. First, in \cite{Brini:2011wi} there has been proposed a mirror curve for torus knots, based on the fact that a torus knot is produced from unknot by an appropriate $SL(2,\mathbb{Z})$ transformation. One may ask how this curve is affected if one substitutes a torus knot by its mirror image. In the construction of \cite{Brini:2011wi}, the mirror curve of a torus knot is obtained from the mirror curve of the unknot via a rational framing transformation. This rational number is determined by the ratio of the two co-prime numbers which define the torus knot. It turns out that the sign of the rational framing transformation determines whether we would find the mirror curve associated with the torus knot or with its mirror. By the rational framing transformation of the unknot mirror curve, we find two distinct curves depending on the sign of the rational framing. One curve corresponds with the torus knot, and the other with the mirror torus knot. 

The second comment concerns links. The identity (\ref{GWrel}) is not restricted to the case of knots, but also holds for Gromov-Witten invariants associated with links. The natural question is that in the case of links, how one realizes the identity (\ref{GWrel}) in the B-model. It was shown in \cite{Aganagic:2013jpa} that for the case of links, instead of mirror curves one needs to work with a higher dimensional variety. This variety coincides with the augmentation variety of the link \cite{Aganagic:2013jpa}. Similar to the case of knots, the augmentation variety of a link can be merely constructed by the knowledge of HOMFLY invariants of the link with $n$ components, colored with $n$ totally symmetric representations. As we saw in the case of knots, this variety can be equivalently constructed in terms of HOMFLY invariants of the link, colored with $n$ totally anti-symmetric representations. Using the statement of the level-rank duality for the case of links and following the same line of argument as for knots, we easily find that (\ref{GWrel}) is fulfilled for the case of links. Furthermore, we find that the augmentation variety of a link is related to the augmentation variety of its mirror by sending $Q\rightarrow Q^{-1}$ and $x_{i}\rightarrow -x_{i}$ ($i=1,2,\cdots,n$). For higher genus and higher hole invariants of links, one would first need to develop a suitable notion of the recursive procedure of \cite{Eynard:2007kz} for higher dimensional varieties. This recursive procedure for higher dimensional varieties is not known yet.  

As the last comment, it is interesting to notice that the level-rank duality holds for composite representations as well. In case of $U(N)$, the most general irreducible representation is specified by a pair of Young tableau \cite{Kioke}. HOMFLY invariants in composite representations are related to amplitudes which are stretched between several Lagrangian branes. Since the level-rank duality holds for composite representations 
\begin{equation*}
{\cal H}^{(\tilde{\cal K})}_{[\mu,\nu]}(Q,q)=(-1)^{|\mu|+|\nu|}\,{\cal H}^{({\cal K})}_{[\mu^{t},\nu^{t}]}(Q^{-1},q)
\end{equation*} 
identity (\ref{GWrel}) is fulfilled for Gromov-Witten invariants associated to stretched amplitudes.

\vspace{.5cm}
\noindent {\bf Acknowledgments.} It is a pleasure to thank Howard Schnitzer for very fruitful discussions on various aspects of the level-rank duality. I would also like to thank Steve Naculich, and Howard Schnitzer for careful comments on this manuscript. This research is supported by NSF FRG grant DMS 1159049 and NSF PHY 1053842.

\vspace{0.5cm}
\noindent {\bf Appendix A.}
\vskip 0.2cm

In this section, we present several open Gromov-Witten invariants associated with the $(3,2)$ and $(5,3)$ torus knots and their mirror images. The left side tables prsent Gromov-Witten invariants of the $(3,2)$ and $(5,3)$ torus knots for a fixed genus and winding vector which has been specified on top of each table. In view of (\ref{GWrel}), the first degree in each table is $d_{\mbox{min}}$ and the last degree represents $d_{\mbox{max}}$. The tables on the right hand side present the corresponding Gromov-Witten invariants for mirror knots. In all these tables, $f$ specifies the framing of the knots.

\newpage

\begin{center}
\begin{tabular}{| c || c |}
\hline
$(3,2)$ & $GW_{(0,2)}(d,\vec{k}=\{2\})$ \\
\hline
$d=0$ & $6 + 3 f + 2 f^2$ \\
\hline
$d=1$ & $-24 - 16 f - 6 f^2$\\
\hline
$d=2$ & $36 + 51/2 f+ 13/2 f^2$\\
\hline
$d=3$ & $-24 - 16 f - 3 f^2$\\
\hline
$d=4$ & $6+ 7/2 f +1/2 f^2$\\
\hline
\end{tabular}
\end{center}

\begin{center}
\begin{tabular}{| c || c |}
\hline
$(3,2)$ & $GW_{(0,3)}(d,\vec{k}=\{3\})$ \\
\hline
$d=0$ & $-36 - 60 f - 65 f^2 - 24 f^3 - 8 f^4$ \\
\hline
$d=1$ & $324 + 540 f + 441 f^2 + 164 f^3 + 36 f^4$\\
\hline
$d=2$ & $-6 (180 + 288 f + 201 f^2 + 68 f^3 + 11 f^4)$\\
\hline
$d=3$ & $3 (600 + 904 f + 559 f^2 + 166 f^3 + 21 f^4)$\\
\hline
$d=4$ & $-1620 - 2268 f - 1251 f^2 - 322 f^3 - 33 f^4$\\
\hline
$d=5$ & $(3 + f)^2 (84 + 52 f + 9 f^2)$\\
\hline
$d=6$ & $-(3 + f)^2 (4 + f)^2$\\
\hline
\end{tabular}
\end{center}

\begin{center}
\begin{tabular}{| c || c |}
\hline
$(3,2)$ & $GW_{(0,3)}(d,\vec{k}=\{2,1\})$ \\
\hline
$d=0$ & $4 (54 + 171 f + 213 f^2 + 165 f^3 + 60 f^4 + 16 f^5)$ \\
\hline
\multirow{2}{*}{$d=1$} & $-2 (1512 + 4068 f + 4614 f^2 + 2943 f^3$ \\ 
& $+ 1000 f^4 + 192 f^5)$\\
\hline
\multirow{2}{*}{$d=2$} & $4 (4014 + 9657 f + 9886 f^2 + 5498 f^3$ \\
& $+ 1665 f^4 + 248 f^5)$\\
\hline
\multirow{2}{*}{$d=3$} & $-2 (22248 + 48828 f + 45286 f^2 + 22383 f^3$\\ 
& $+ 5960 f^4 + 720 f^5)$\\
\hline
\multirow{2}{*}{$d=4$} & $2 (36180 + 73026 f + 61575 f^2 + 27230 f^3$ \\
& $+ 6365 f^4 + 642 f^5)$\\
\hline
\multirow{2}{*}{$d=5$} & $-4 (17964 + 33438 f + 25673 f^2 + 10184 f^3$ \\
& $+ 2095 f^4 + 180 f^5)$\\
\hline
\multirow{2}{*}{$d=6$} & $4 (10746 + 18447 f + 12898 f^2+ 4595 f^3 $\\ 
& $+ 835 f^4 + 62 f^5)$\\
\hline
$d=7$ & $-4 (3 + f)^2 (396 + 362 f + 113 f^2 + 12 f^3)$\\
\hline
$d=8$ & $2 (3 + f)^2 (4 + f)^2 (7 + 2 f)$\\
\hline
\end{tabular}
\end{center}

\begin{center}
\begin{tabular}{| c || c |}
\hline
$(3,2)$ & $GW_{(0,4)}(d,\vec{k}=\{4\})$ \\
\hline
\multirow{2}{*}{$d=0$} & $2 (216 + 648 f + 1026 f^2 + 852 f^3$ \\
& $ + 495 f^4 + 144 f^5 + 32 f^6)$ \\
\hline
\multirow{2}{*}{$d=1$} & $-3 (2160 + 6048 f + 8136 f^2 + 6152 f^3$ \\ 
&  $+ 2943 f^4 + 800 f^5 + 128 f^6)$ \\ 
\hline
\multirow{2}{*}{$d=2$} & $4 (9180 + 24084 f + 28971 f^2 + 19772 f^3 $ \\
& $+ 8247 f^4 + 1998 f^5 + 248 f^6)$\\
\hline
\multirow{2}{*}{$d=3$} & $-108432 - 266976 f - 292968 f^2 - 181144 f^3$\\ 
& $- 67149 f^4 -14304 f^5 - 1440 f^6$\\
\hline
\multirow{2}{*}{$d=4$} & $6 (31320 + 72360 f + 73026 f^2 + 41050 f^3$ \\
& $+ 13615 f^4 + 2546 f^5 + 214 f^6)$\\
\hline
\multirow{2}{*}{$d=5$} & $-8 (24894 + 53892 f + 50157 f^2 + 25673 f^3$ \\
& $+ 7638 f^4 + 1257 f^5 + 90 f^6)$\\
\hline
\multirow{2}{*}{$d=6$} & $2 (63720 + 128952 f + 110682 f^2 + 51592 f^3$\\ 
& $+ 13785 f^4 + 2004 f^5 + 124 f^6)$\\
\hline
$d=7$ & $-24 (3 + f)^3 (7 + 2 f) (10 + 6 f + f^2)$\\
\hline
$d=8$ & $4 (3 + f)^3 (4 + f)^3$\\
\hline
\end{tabular}
\end{center}

\begin{center}
\begin{tabular}{| c || c |}
\hline
$\tilde{(3,2)}$ & $GW_{(0,2)}(d,\vec{k}=\{2\})$ \\
\hline
$d=0$ & $6 + 7/2 f + 1/2 f^2$ \\
\hline
$d=1$ & $-24 - 16 f - 3 f^2$\\
\hline
$d=2$ & $36 + 51/2 f + 13/2 f^2$\\
\hline
$d=3$ & $-24 - 16 f - 6 f^2$\\
\hline
$d=4$ & $6 + 3 f + 2 f^2$\\
\hline
\end{tabular}
\end{center}

\begin{center}
\begin{tabular}{| c || c |}
\hline
$\widetilde{(3,2)}$ & $GW_{(0,3)}(d,\vec{k}=\{3\})$ \\
\hline
$d=0$ & $(3 + f)^2 (4 + f)^2$ \\
\hline
$d=1$ & $-(3 + f)^2 (84 + 52 f + 9 f^2)$\\
\hline
$d=2$ & $1620 + 2268 f + 1251 f^2 + 322 f^3 + 33 f^4$\\
\hline
$d=3$ & $-3 (600 + 904 f + 559 f^2 + 166 f^3 + 21 f^4)$\\
\hline
$d=4$ & $6 (180 + 288 f + 201 f^2 + 68 f^3 + 11 f^4)$\\
\hline
$d=5$ & $-324 - 540 f - 441 f^2 - 164 f^3 - 36 f^4$\\
\hline
$d=6$ & $36 + 60 f + 65 f^2 + 24 f^3 + 8 f^4$\\
\hline
\end{tabular}
\end{center}

\begin{center}
\begin{tabular}{| c || c |}
\hline
$\widetilde{(3,2)}$ & $GW_{(0,3)}(d,\vec{k}=\{2,1\})$ \\
\hline
$d=0$ & $-2 (3 + f)^2 (4 + f)^2 (7 + 2 f)$ \\
\hline
$d=1$ & $4 (3 + f)^2 (396 + 362 f + 113 f^2 + 12 f^3)$ \\ 
\hline
\multirow{2}{*}{$d=2$} & $-4 (10746 + 18447 f + 12898 f^2 + 4595 f^3$ \\
& $+ 835 f^4 + 62 f^5)$\\
\hline
\multirow{2}{*}{$d=3$} & $4 (17964 + 33438 f + 25673 f^2 + 10184 f^3$\\ 
& $+ 2095 f^4 + 180 f^5)$\\
\hline
\multirow{2}{*}{$d=4$} & $-2 (36180 + 73026 f + 61575 f^2 + 27230 f^3$ \\
& $+ 6365 f^4 + 642 f^5)$\\
\hline
\multirow{2}{*}{$d=5$} & $2 (22248 + 48828 f + 45286 f^2 + 22383 f^3$ \\
& $ + 5960 f^4 + 720 f^5)$\\
\hline
\multirow{2}{*}{$d=6$} & $-4 (4014 + 9657 f + 9886 f^2 + 5498 f^3$\\ 
& $ + 1665 f^4 + 248 f^5)$\\
\hline
\multirow{2}{*}{$d=7$} & $2 (1512 + 4068 f + 4614 f^2 + 2943 f^3$\\
& $ + 1000 f^4 + 192 f^5)$\\
\hline
$d=8$ & $-4 (54 + 171 f + 213 f^2 + 165 f^3 + 60 f^4 + 16 f^5)$\\
\hline
\end{tabular}
\end{center}

\begin{center}
\begin{tabular}{| c || c |}
\hline
$\widetilde{(3,2)}$ & $GW_{(0,4)}(d,\vec{k}=\{4\})$ \\
\hline
$d=0$ & $4 (3 + f)^3 (4 + f)^3$ \\
\hline
$d=1$ & $-24 (3 + f)^3 (7 + 2 f) (10 + 6 f + f^2)$ \\ 
\hline
\multirow{2}{*}{$d=2$} & $2 (63720 + 128952 f + 110682 f^2 + 51592 f^3$ \\
& $+ 13785 f^4 + 2004 f^5 +124 f^6)$\\
\hline
\multirow{2}{*}{$d=3$} & $-8 (24894 + 53892 f + 50157 f^2 + 25673 f^3$\\ 
& $+ 7638 f^4 + 1257 f^5 + 90 f^6)$\\
\hline
\multirow{2}{*}{$d=4$} & $6 (31320 + 72360 f + 73026 f^2 + 41050 f^3$ \\
& $ + 13615 f^4 + 2546 f^5 + 214 f^6)$\\
\hline
\multirow{2}{*}{$d=5$} & $-108432 - 266976 f - 292968 f^2 - 181144 f^3$ \\
& $ - 67149 f^4 - 14304 f^5 - 1440 f^6$\\
\hline
\multirow{2}{*}{$d=6$} & $4 (9180 + 24084 f + 28971 f^2 + 19772 f^3$\\ 
& $+ 8247 f^4 + 1998 f^5 + 248 f^6)$\\
\hline
\multirow{2}{*}{$d=7$} & $-3 (2160 + 6048 f + 8136 f^2 + 6152 f^3$\\
& $ + 2943 f^4 + 800 f^5 + 128 f^6)$\\
\hline
\multirow{2}{*}{$d=8$} & $2 (216 + 648 f + 1026 f^2 + 852 f^3$\\
& $ + 495 f^4 + 144 f^5 + 32 f^6)$\\
\hline
\end{tabular}
\end{center}


\newpage

\begin{center}
\begin{tabular}{| c || c |}
\hline
$(3,2)$ & $GW_{(1,2)}(d,\vec{k}=\{2\})$ \\
\hline
$d=0$ & $1/6(30 + 27 f + 29 f^2 + 3 f^3 + f^4)$ \\
\hline
$d=1$ & $1/6(-156 - 176 f - 99 f^2 - 16 f^3 - 3 f^4)$\\
\hline
$d=2$ & $1/24(1008 + 1146 f + 515 f^2 + 102 f^3 + 13 f^4)$\\
\hline
$d=3$ & $1/12(-312 - 352 f - 153 f^2 - 32 f^3 - 3 f^4)$\\
\hline
$d=4$ & $1/24(2 + f) (3 + f) (4 + f) (5 + f)$\\
\hline
\end{tabular}
\end{center}

\begin{center}
\begin{tabular}{| c || c |}
\hline
$(3,2)$ & $GW_{(1,3)}(d,\vec{k}=\{3\})$ \\
\hline
\multirow{2}{*}{$d=0$} & $1/24(-1404 - 3780 f - 5391 f^2 - 3288 f^3$\\
& $ - 1722 f^4 - 288 f^5 - 64 f^6)$ \\
\hline
\multirow{2}{*}{$d=1$} & $1/24(17820 + 42660 f + 47367 f^2 + 28028 f^3$\\
& $ + 10566 f^4 + 1968 f^5 + 288 f^6)$\\
\hline
\multirow{2}{*}{$d=2$} & $1/4(-12060 - 26640 f - 25995 f^2 - 13972 f^3$\\
& $ - 4485 f^4 - 816 f^5 - 88 f^6)$\\
\hline
\multirow{2}{*}{$d=3$} & $3/8(15192 + 31368 f + 27939 f^2 + 13702 f^3$\\
& $+ 3965 f^4 + 664 f^5 + 56 f^6)$\\
\hline
\multirow{2}{*}{$d=4$} & $1/24(-133164 - 259524 f - 215037 f^2 - 97462 f^3$\\
& $ - 25803 f^4 - 3864 f^5 - 264 f^6)$\\
\hline
\multirow{2}{*}{$d=5$} & $1/24(3 + f)^2 (7212 + 8620 f+ 3951 f^2$\\
& $ + 840 f^3 + 72 f^4)$\\
\hline
$d=6$ & $-1/24(3 + f)^2 (4 + f)^2 (87 + 56 f + 8 f^2)$\\
\hline
\end{tabular}
\end{center}

\begin{center}
\begin{tabular}{| c || c |}
\hline
$(3,2)$ & $GW_{(1,3)}(d,\vec{k}=\{2,1\})$ \\
\hline
\multirow{2}{*}{$d=0$} & $1/3(2592 + 10584 f + 18366 f^2 + 19383 f^3$ \\
& $+ 11610 f^4 + 5185 f^5 + 1092 f^6 + 208 f^7)$\\
\hline
\multirow{3}{*}{$d=1$} & $1/6(-94392 - 327276 f - 499818 f^2 $ \\ 
& $- 444273 f^3- 243750 f^4 - 88197 f^5$\\
& $- 18200 f^6 - 2496 f^7)$\\
\hline
\multirow{2}{*}{$d=2$} & $1/3(303768 + 941652 f + 1291552 f^2$ \\
& $+ 1021558 f^3+ 504805 f^4 + 159572 f^5$\\
& $ + 30303 f^6 + 3224 f^7)$\\
\hline
\multirow{3}{*}{$d=3$} & $1/6(-1960632 - 5564964 f - 6968098 f^2 $\\ 
& $- 5006001 f^3-2242510 f^4 - 635877 f^5$\\
& $- 108472 f^6 - 9360 f^7)$\\
\hline
\multirow{3}{*}{$d=4$} & $1/6(3620160 + 9530496 f + 11012736 f^2$ \\
& $+ 7263475 f^3 + 2971305 f^4 + 762423 f^5$\\
& $+ 115843 f^6 + 8346 f^7)$ \\
\hline
\multirow{3}{*}{$d=5$} & $-1/3(2007108 + 4940202 f + 5305967 f^2$ \\
& $ + 3233194 f^3 + 1213040 f^4 + 282288 f^5 $\\
& $+ 38129 f^6 + 2340 f^7)$\\
\hline
\multirow{3}{*}{$d=6$} & $1/3(1326024 + 3068268 f + 3077632 f^2$\\ 
& $+ 1738636 f^3 + 599215 f^4 + 126458 f^5$\\
& $ + 15197 f^6 + 806 f^7)$\\
\hline
\multirow{2}{*}{$d=7$} & $-1/3(3 + f)^2 (53604 + 81334 f + 49803 f^2$\\
& $+ 15438 f^3 + 2431 f^4 + 156 f^5)$\\
\hline
\multirow{2}{*}{$d=8$} & $1/6(3 + f)^2 (4 + f)^2 (7 + 2 f)$\\
& $(148 + 91 f + 13 f^2)$\\
\hline
\end{tabular}
\end{center}

\begin{center}
\begin{tabular}{| c || c |}
\hline
$\widetilde{(3,2)}$ & $GW_{(1,2)}(d,\vec{k}=\{2\})$ \\
\hline
$d=0$ & $1/24(2 + f) (3 + f) (4 + f) (5 + f)$ \\
\hline
$d=1$ & $1/12(-312 - 352 f - 153 f^2 - 32 f^3 - 3 f^4)$\\
\hline
$d=2$ & $1/24(1008 + 1146 f + 515 f^2 + 102 f^3 + 13 f^4)$\\
\hline
$d=3$ & $1/6(-156 - 176 f - 99 f^2 - 16 f^3 - 3 f^4)$\\
\hline
$d=4$ & $1/6(30 + 27 f + 29 f^2 + 3 f^3 + f^4)$\\
\hline
\end{tabular}
\end{center}

\begin{center}
\begin{tabular}{| c || c |}
\hline
$\widetilde{(3,2)}$ & $GW_{(1,3)}(d,\vec{k}=\{3\})$ \\
\hline
$d=0$ & $1/24(3 + f)^2 (4 + f)^2 (87 + 56 f + 8 f^2)$ \\
\hline
\multirow{2}{*}{$d=1$} & $-1/24(3 + f)^2 (7212 + 8620 f + 3951 f^2$\\
& $+ 840 f^3 + 72 f^4)$\\
\hline
\multirow{2}{*}{$d=2$} & $1/24(133164 + 259524 f + 215037 f^2 + 97462 f^3$\\
& $+ 25803 f^4 + 3864 f^5 + 264 f^6)$\\
\hline
\multirow{2}{*}{$d=3$} & $-3/8(15192 + 31368 f + 27939 f^2 + 13702 f^3$\\
& $+ 3965 f^4 + 664 f^5 + 56 f^6)$\\
\hline
\multirow{2}{*}{$d=4$} & $1/4(12060 + 26640 f + 25995 f^2 + 13972 f^3$\\
& $+ 4485 f^4 + 816 f^5 + 88 f^6)$\\
\hline
\multirow{2}{*}{$d=5$} & $1/24(-17820 - 42660 f - 47367 f^2 - 28028 f^3$\\
& $- 10566 f^4 - 1968 f^5 - 288 f^6)$\\
\hline
\multirow{2}{*}{$d=6$} & $1/24(1404 + 3780 f + 5391 f^2 + 3288 f^3$\\
& $+ 1722 f^4 + 288 f^5 + 64 f^6)$\\
\hline
\end{tabular}
\end{center}

\begin{center}
\begin{tabular}{| c || c |}
\hline
$\widetilde{(3,2)}$ & $GW_{(1,3)}(d,\vec{k}=\{2,1\})$ \\
\hline
\multirow{2}{*}{$d=0$} & $-1/6(3 + f)^2 (4 + f)^2 (7 + 2 f)$ \\
& $(148 + 91 f + 13 f^2)$\\
\hline
\multirow{2}{*}{$d=1$} & $1/3(3 + f)^2 (53604 + 81334 f + 49803 f^2$ \\
& $+ 15438 f^3 + 2431 f^4 + 156 f^5)$\\ 
\hline
\multirow{3}{*}{$d=2$} & $-1/3(1326024 + 3068268 f + 3077632 f^2$ \\
& $+ 1738636 f^3 + 599215 f^4 + 126458 f^5$\\
& $ + 15197 f^6 + 806 f^7)$\\
\hline
\multirow{3}{*}{$d=3$} & $1/3(2007108 + 4940202 f + 5305967 f^2$\\ 
& $+ 3233194 f^3 + 1213040 f^4 + 282288 f^5$\\
& $ + 38129 f^6 + 2340 f^7)$\\
\hline
\multirow{3}{*}{$d=4$} & $1/6(-3620160 - 9530496 f - 11012736 f^2$ \\
& $- 7263475 f^3 - 2971305 f^4 - 762423 f^5 $\\
& $- 115843 f^6 - 8346 f^7)$\\
\hline
\multirow{3}{*}{$d=5$} & $1/6(1960632 + 5564964 f + 6968098 f^2$ \\
& $+ 5006001 f^3 + 2242510 f^4 + 635877 f^5$\\
& $+ 108472 f^6 + 9360 f^7)$\\
\hline
\multirow{3}{*}{$d=6$} & $-1/3(303768 + 941652 f + 1291552 f^2$\\ 
& $ + 1021558 f^3+ 504805 f^4 + 159572 f^5$\\
& $+ 30303 f^6 + 3224 f^7)$\\
\hline
\multirow{3}{*}{$d=7$} & $1/6(94392 + 327276 f + 499818 f^2$\\
& $ + 444273 f^3+ 243750 f^4 + 88197 f^5$\\
& $ + 18200 f^6 + 2496 f^7)$\\
\hline
\multirow{2}{*}{$d=8$} & $-1/3(2592 + 10584 f + 18366 f^2 + 19383 f^3$\\
& $+ 11610 f^4 + 5185 f^5 + 1092 f^6 + 208 f^7)$\\
\hline
\end{tabular}
\end{center}


\begin{center}
\begin{tabular}{| c || c |}
\hline
$(3,2)$ & $GW_{(2,2)}(d,\vec{k}=\{2\})$ \\
\hline
\multirow{2}{*}{$d=0$} & $1/360(510 + 951 f + 1083 f^2 + 270 f^3$ \\
& $+ 145 f^4 + 9 f^5 + 2 f^6)$\\
\hline
\multirow{2}{*}{$d=1$} & $1/360(-4020 - 6712 f - 5049 f^2 - 1760 f^3$\\
& $ - 495 f^4 - 48 f^5 - 6 f^6)$\\
\hline
\multirow{2}{*}{$d=2$} & $1/1440(28080 + 45594 f + 30999 f^2 + 11460 f^3$\\
& $ + 2575 f^4 + 306 f^5 + 26 f^6)$\\
\hline
\multirow{2}{*}{$d=3$} & $1/720(-8040 - 13424 f - 9369 f^2 - 3520 f^3$\\
& $- 765 f^4 - 96 f^5 - 6 f^6)$\\
\hline
\multirow{2}{*}{$d=4$} & $1/1440(2 + f) (3 + f) (4 + f) (5 + f)$\\
& $ (17 + 14 f + 2 f^2)$\\
\hline
\end{tabular}
\end{center}

\begin{center}
\begin{tabular}{| c || c |}
\hline
$(3,2)$ & $GW_{(2,3)}(d,\vec{k}=\{3\})$ \\
\hline
\multirow{3}{*}{$d=0$} & $1/1920(-82836 - 313740 f - 557325 f^2$\\
& $- 526872 f^3 - 363916 f^4 - 122880 f^5 $ \\
& $ - 42800 f^6 - 4992 f^7 - 832 f^8)$\\
\hline
\multirow{3}{*}{$d=1$} & $1/1920(1523124 + 4810860 f + 6913989 f^2$\\
& $ + 5709652 f^3 + 3060072 f^4 + 1043360 f^5$\\
& $+ 260592 f^6 + 34112 f^7 + 3744 f^8$\\
\hline
\multirow{3}{*}{$d=2$} & $-1/320(1279620 + 3643200 f + 4645869 f^2$\\
& $ + 3443844 f^3 + 1638167 f^4 + 517408 f^5 $\\
& $+ 109952 f^6 + 14144 f^7 + 1144 f^8)$\\
\hline
\multirow{3}{*}{$d=3$} & $1/640(5547960 + 14702760 f + 17315091 f^2$\\
& $+ 11856318 f^3 + 5201617 f^4 + 1514064 f^5$\\
& $+ 290368 f^6 + 34528 f^7 + 2184 f^8)$\\
\hline
\multirow{3}{*}{$d=4$} & $1/1920(-17646660 - 44554860 f - 49692519 f^2$\\
& $ - 32095466 f^3 - 13204461 f^4 - 3571888 f^5$\\
& $ - 628032 f^6 - 66976 f^7 - 3432 f^8)$\\
\hline
\multirow{3}{*}{$d=5$} & $1/1920(3 + f)^2 (994884 + 1786724 f$\\
& $ + 1343205 f^2 + 544816 f^3 + 127128 f^4$\\
& $+ 16432 f^5 + 936 f^6)$\\
\hline
\multirow{2}{*}{$d=6$} & $-1/1920(3 + f)^2 (4 + f)^2$\\
& $(11901 + 15568 f + 7320 f^2 + 1456 f^3 + 104 f^4)$\\
\hline
\end{tabular}
\end{center}

\begin{center}
\begin{tabular}{| c || c |}
\hline
$(5,3)$ & $GW_{(0,2)}(d,\vec{k}=\{2\})$ \\
\hline
$d=0$ & $945 + 469/2 f + 49/2 f^2$ \\
\hline
$d=1$ & $-4620 - 1148 f - 105 f^2$\\
\hline
$d=2$ & $9225 + 4515/2 f + 365/2 f^2$\\
\hline
$d=3$ & $-9600 - 2280 f - 164 f^2$\\
\hline
$d=4$ & $5475 + 1245 f + 80 f^2$\\
\hline
$d=5$ & $-1620 - 348 f - 20 f^2$\\
\hline
$d=6$ & $195 + 39 f + 2 f^2$\\
\hline
\end{tabular}
\end{center}

\begin{center}
\begin{tabular}{| c || c |}
\hline
$\widetilde{(3,2)}$ & $GW_{(2,2)}(d,\vec{k}=\{2\})$ \\
\hline
\multirow{2}{*}{$d=0$} & $1/1440(2 + f) (3 + f) (4 + f) (5 + f)$ \\
& $ (17 + 14 f + 2 f^2)$\\
\hline
\multirow{2}{*}{$d=1$} & $1/720(-8040 - 13424 f - 9369 f^2 - 3520 f^3$\\
& $- 765 f^4 - 96 f^5 - 6 f^6)$\\
\hline
\multirow{2}{*}{$d=2$} & $1/1440(28080 + 45594 f + 30999 f^2 + 11460 f^3$\\
& $ + 2575 f^4 + 306 f^5 + 26 f^6)$\\
\hline
\multirow{2}{*}{$d=3$} & $1/360(-4020 - 6712 f - 5049 f^2 - 1760 f^3$\\
& $- 495 f^4 - 48 f^5 - 6 f^6)$\\
\hline
\multirow{2}{*}{$d=4$} & $1/360(510 + 951 f + 1083 f^2 + 270 f^3$\\
& $ + 145 f^4 + 9 f^5 + 2 f^6)$\\
\hline
\end{tabular}
\end{center}

\vskip 1.7cm

\begin{center}
\begin{tabular}{| c || c |}
\hline
$\widetilde{(3,2)}$ & $GW_{(2,3)}(d,\vec{k}=\{3\})$ \\
\hline
\multirow{2}{*}{$d=0$} & $1/1920(3 + f)^2 (4 + f)^2$ \\
& $(11901 + 15568 f + 7320 f^2 + 1456 f^3 + 104 f^4)$\\
\hline
\multirow{3}{*}{$d=1$} & $-1/1920(3 + f)^2 (994884 + 1786724 f$\\
& $ + 1343205 f^2 + 544816 f^3 + 127128 f^4$\\
& $+ 16432 f^5 + 936 f^6)$\\
\hline
\multirow{3}{*}{$d=2$} & $1/1920(17646660 + 44554860 f + 49692519 f^2$\\
& $+ 32095466 f^3 + 13204461 f^4 + 3571888 f^5$\\
& $ + 628032 f^6 + 66976 f^7 + 3432 f^8)$\\
\hline
\multirow{3}{*}{$d=3$} & $-1/640(5547960 + 14702760 f + 17315091 f^2$\\
& $+ 11856318 f^3 + 5201617 f^4 + 1514064 f^5$\\
& $+ 290368 f^6 + 34528 f^7 + 2184 f^8)$\\
\hline
\multirow{3}{*}{$d=4$} & $1/320(1279620 + 3643200 f + 4645869 f^2$\\
& $+ 3443844 f^3 + 1638167 f^4 + 517408 f^5$\\
& $+ 109952 f^6 + 14144 f^7 + 1144 f^8)$\\
\hline
\multirow{3}{*}{$d=5$} & $1/1920(-1523124 - 4810860 f - 6913989 f^2$\\
& $- 5709652 f^3 - 3060072 f^4 - 1043360 f^5$\\
& $- 260592 f^6 - 34112 f^7 - 3744 f^8)$\\
\hline
\multirow{3}{*}{$d=6$} & $1/1920(82836 + 313740 f + 557325 f^2$\\
& $+ 526872 f^3 + 363916 f^4 + 122880 f^5$\\
& $+ 42800 f^6 + 4992 f^7 + 832 f^8)$\\
\hline
\end{tabular}
\end{center}

\vskip 1.74cm

\begin{center}
\begin{tabular}{| c || c |}
\hline
$\widetilde{(5,3)}$ & $GW_{(0,2)}(d,\vec{k}=\{2\})$ \\
\hline
$d=0$ & $195 + 39 f + 2 f^2$ \\
\hline
$d=1$ & $-1620 - 348 f - 20 f^2$\\
\hline
$d=2$ & $5475 + 1245 f + 80 f^2$\\
\hline
$d=3$ & $-9600 - 2280 f - 164 f^2$\\
\hline
$d=4$ & $9225 + 4515/2 f + 365/2 f^2$\\
\hline
$d=5$ & $-4620 - 1148 f - 105 f^2$\\
\hline
$d=6$ & $945 + 469/2 f + 49/2 f^2$\\
\hline
\end{tabular}
\end{center}

\newpage

\begin{center}
\begin{tabular}{| c || c |}
\hline
$(5,3)$ & $GW_{(0,3)}(d,\vec{k}=\{3\})$ \\
\hline
\multirow{2}{*}{$d=0$} & $-446175 - 244710 f - 57598 f^2$\\
& $- 6566 f^3 - 343 f^4$ \\
\hline
\multirow{2}{*}{$d=1$} & $3526875 + 1894050 f + 426015 f^2$\\
& $ + 46214 f^3 + 2205 f^4$\\
\hline
\multirow{2}{*}{$d=2$} & $-35 (348300 + 182520 f + 39258 f^2$\\
& $+ 4042 f^3 + 177 f^4)$\\
\hline
\multirow{2}{*}{$d=3$} & $3 (8058000 + 4106900 f + 844565 f^2$\\
& $ + 82362 f^3 + 3323 f^4)$\\
\hline
\multirow{2}{*}{$d=4$} & $-2 (15148350 + 7484670 f + 1470312 f^2$\\
& $+ 135572 f^3 + 5055 f^4)$\\
\hline
\multirow{2}{*}{$d=5$} & $6 (4146300 + 1979460 f + 370911 f^2$\\
& $ + 32284 f^3 + 1115 f^4)$\\
\hline
\multirow{2}{*}{$d=6$} & $-4 (3345750 + 1537950 f + 274345 f^2$\\
& $ + 22503 f^3 + 721 f^4)$\\
\hline
\multirow{2}{*}{$d=7$} & $60 (75825 + 33435 f + 5664 f^2$\\
& $ + 437 f^3 + 13 f^4)$\\
\hline
\multirow{2}{*}{$d=8$} & $-3 (295875 + 124650 f + 19995 f^2$\\
& $  + 1448 f^3 + 40 f^4)$\\
\hline
\multirow{2}{*}{$d=9$} & $75825 + 30390 f + 4601 f^2$\\
& $ + 312 f^3 + 8 f^4$\\
\hline
\end{tabular}
\end{center}

\begin{center}
\begin{tabular}{| c || c |}
\hline
$(5,3)$ & $GW_{(0,3)}(d,\vec{k}=\{2,1\})$ \\
\hline
\multirow{2}{*}{$d=0$} & $14 (4224375 + 3215595 f + 1035380 f^2$ \\
& $+177757 f^3 + 16415 f^4 + 686 f^5)$\\
\hline
\multirow{2}{*}{$d=1$} & $-2 (317398500 + 234323100 f + 72838140 f^2$ \\ 
& $+ 11992553 f^3 + 1054970 f^4 + 41160 f^5)$\\
\hline
\multirow{2}{*}{$d=2$} & $2 (1538574975 + 1101566115 f + 330485465 f^2$ \\
& $ + 52191331 f^3 + 4372900 f^4 + 159740 f^5)$\\
\hline
\multirow{2}{*}{$d=3$} & $-2 (4454199000 + 3091824900 f + 894962550 f^2$\\ 
& $+ 135559503 f^3 + 10817310 f^4 + 370888 f^5)$\\
\hline
\multirow{2}{*}{$d=4$} & $2 (8577231750 + 5769410400 f + 1610539545 f^2$ \\
& $+ 233934894 f^3 + 17778565 f^4 + 573330 f^5)$\\
\hline
\multirow{2}{*}{$d=5$} & $-4 (5786823600 + 3769386870 f + 1014168265 f^2$ \\
& $+ 141218952 f^3 + 10221110 f^4 + 310560 f^5)$\\
\hline
\multirow{2}{*}{$d=6$} & $8 (2805132375 + 1767920925 f + 458136460 f^2$\\ 
& $+ 61128324 f^3 + 4213285 f^4 + 120788 f^5)$\\
\hline
\multirow{2}{*}{$d=7$} & $-4 (3937572000 + 2398708710 f + 598186347 f^2$\\
& $+ 76436100 f^3 + 5016310 f^4 + 135840 f^5)$\\
\hline
\multirow{2}{*}{$d=8$} & $2 (3971581875 + 2335831875 f + 560015835 f^2$\\
& $+ 68480901 f^3 + 4278090 f^4 + 109520 f^5)$\\
\hline
\multirow{2}{*}{$d=9$} & $-4 (701934750 + 398037900 f + 91641805 f^2$\\
& $ + 10715313 f^3 + 636960 f^4 + 15424 f^5)$\\
\hline
\multirow{2}{*}{$d=10$} & $6 (110087775 + 60099675 f + 13270981 f^2$\\
& $+ 1482297 f^3 + 83800 f^4 + 1920 f^5)$\\
\hline
\multirow{2}{*}{$d=11$} & $-8 (11612250 + 6093300 f + 1288675 f^2$\\
& $+ 137349 f^3 + 7380 f^4 + 160 f^5)$\\
\hline
\multirow{2}{*}{$d=12$} & $4 (1477575 + 743940 f + 150468 f^2$\\
& $+ 15285 f^3 + 780 f^4 + 16 f^5)$\\
\hline
\end{tabular}
\end{center}

\begin{center}
\begin{tabular}{| c || c |}
\hline
$\widetilde{(5,3)}$ & $GW_{(0,3)}(d,\vec{k}=\{3\})$ \\
\hline
\multirow{2}{*}{$d=0$} & $-75825 - 30390 f - 4601 f^2$ \\
& $- 312 f^3 - 8 f^4$\\
\hline
\multirow{2}{*}{$d=1$} & $3 (295875 + 124650 f + 19995 f^2$\\
& $ + 1448 f^3 + 40 f^4)$\\
\hline
\multirow{2}{*}{$d=2$} & $-60 (75825 + 33435 f + 5664 f^2$\\
& $+ 437 f^3 + 13 f^4)$\\
\hline
\multirow{2}{*}{$d=3$} & $4 (3345750 + 1537950 f + 274345 f^2$\\
& $+ 22503 f^3 + 721 f^4)$\\
\hline
\multirow{2}{*}{$d=4$} & $-6 (4146300 + 1979460 f + 370911 f^2$\\
& $+ 32284 f^3 + 1115 f^4)$\\
\hline
\multirow{2}{*}{$d=5$} & $2 (15148350 + 7484670 f + 1470312 f^2$\\
& $ + 135572 f^3 + 5055 f^4)$\\
\hline
\multirow{2}{*}{$d=6$} & $-3 (8058000 + 4106900 f + 844565 f^2$\\
& $+ 82362 f^3 + 3323 f^4)$\\
\hline
\multirow{2}{*}{$d=7$} & $35 (348300 + 182520 f + 39258 f^2$\\
& $+ 4042 f^3 + 177 f^4)$\\
\hline
\multirow{2}{*}{$d=8$} & $-3526875 - 1894050 f - 426015 f^2$\\
& $ - 46214 f^3 - 2205 f^4$\\
\hline
\multirow{2}{*}{$d=9$} & $446175 + 244710 f + 57598 f^2$\\
& $+ 6566 f^3 + 343 f^4$\\
\hline
\end{tabular}
\end{center}

\begin{center}
\begin{tabular}{| c || c |}
\hline
$\widetilde{(5,3)}$ & $GW_{(0,3)}(d,\vec{k}=\{2,1\})$ \\
\hline
\multirow{2}{*}{$d=0$} & $-4 (1477575 + 743940 f + 150468 f^2$\\
& $+ 15285 f^3 + 780 f^4 + 16 f^5)$\\
\hline
\multirow{2}{*}{$d=1$} & $8 (11612250 + 6093300 f + 1288675 f^2$\\
& $+ 137349 f^3 + 7380 f^4 + 160 f^5)$\\
\hline
\multirow{2}{*}{$d=2$} & $-6 (110087775 + 60099675 f + 13270981 f^2$\\
& $+ 1482297 f^3 + 83800 f^4 + 1920 f^5)$\\
\hline
\multirow{2}{*}{$d=3$} & $4 (701934750 + 398037900 f + 91641805 f^2$\\
& $ + 10715313 f^3 + 636960 f^4 + 15424 f^5)$\\
\hline
\multirow{2}{*}{$d=4$} & $-2 (3971581875 + 2335831875 f + 560015835 f^2$\\
& $+ 68480901 f^3 + 4278090 f^4 + 109520 f^5)$\\
\hline
\multirow{2}{*}{$d=5$} & $4 (3937572000 + 2398708710 f + 598186347 f^2$\\
& $+ 76436100 f^3 + 5016310 f^4 + 135840 f^5)$\\
\hline
\multirow{2}{*}{$d=6$} & $-8 (2805132375 + 1767920925 f + 458136460 f^2$\\ 
& $+ 61128324 f^3 + 4213285 f^4 + 120788 f^5)$\\
\hline
\multirow{2}{*}{$d=7$} & $4 (5786823600 + 3769386870 f + 1014168265 f^2$ \\
& $+ 141218952 f^3 + 10221110 f^4 + 310560 f^5)$\\
\hline
\multirow{2}{*}{$d=8$} & $-2 (8577231750 + 5769410400 f + 1610539545 f^2$ \\
& $+ 233934894 f^3 + 17778565 f^4 + 573330 f^5)$\\
\hline
\multirow{2}{*}{$d=9$} & $2 (4454199000 + 3091824900 f + 894962550 f^2$\\ 
& $+ 135559503 f^3 + 10817310 f^4 + 370888 f^5)$\\
\hline
\multirow{2}{*}{$d=10$} & $-2 (1538574975 + 1101566115 f + 330485465 f^2$ \\
& $ + 52191331 f^3 + 4372900 f^4 + 159740 f^5)$\\
\hline
\multirow{2}{*}{$d=11$} & $2 (317398500 + 234323100 f + 72838140 f^2$ \\ 
& $+ 11992553 f^3 + 1054970 f^4 + 41160 f^5)$\\
\hline
\multirow{2}{*}{$d=12$} & $-14 (4224375 + 3215595 f + 1035380 f^2$ \\
& $+177757 f^3 + 16415 f^4 + 686 f^5)$\\
\hline
\end{tabular}
\end{center}


\newpage

\begin{center}
\begin{tabular}{| c || c |}
\hline
$(5,3)$ & $GW_{(1,2)}(d,\vec{k}=\{2\})$ \\
\hline
\multirow{2}{*}{$d=0$} & $7/24(35640 + 12934 f + 2117 f^2$ \\
& $ + 134 f^3 + 7 f^4)$\\
\hline
\multirow{2}{*}{$d=1$} & $-7/12(87420 + 31868 f + 4905 f^2$\\
& $ + 328 f^3 + 15 f^4)$\\
\hline
\multirow{2}{*}{$d=2$} & $5/24(484200 + 176190 f + 26099 f^2$\\
& $ + 1806 f^3 + 73 f^4)$\\
\hline
\multirow{2}{*}{$d=3$} & $1/3(-308400 - 111870 f - 16231 f^2$\\
& $- 1140 f^3 - 41 f^4)$\\
\hline
\multirow{2}{*}{$d=4$} & $5/6(68250 + 24750 f + 3559 f^2$\\
& $+ 249 f^3 + 8 f^4)$\\
\hline
\multirow{2}{*}{$d=5$} & $1/3(-48735 - 17769 f - 2545 f^2$\\
& $ - 174 f^3 - 5 f^4)$\\
\hline
\multirow{2}{*}{$d=6$} & $1/6(11370 + 4194 f + 596 f^2$\\
& $+ 39 f^3 + f^4)$\\
\hline
\end{tabular}
\end{center}

\begin{center}
\begin{tabular}{| c || c |}
\hline
$(5,3)$ & $GW_{(1,3)}(d,\vec{k}=\{3\})$ \\
\hline
\multirow{3}{*}{$d=0$} & $1/24(-289459575 - 207901890 f$\\
& $- 66413862 f^2 - 11810810 f^3 - 1268583 f^4$ \\
& $- 78792 f^5 - 2744 f^6)$\\
\hline
\multirow{3}{*}{$d=1$} & $1/24(2375206875 + 1671279750 f$\\
& $+ 517317525 f^2 + 89166410 f^3 + 9219015 f^4$\\
& $ + 554568 f^5 + 17640 f^6)$\\
\hline
\multirow{3}{*}{$d=2$} & $-35/24 (242441100 + 167036040 f$\\
& $ + 50197806 f^2 + 8389630 f^3 + 836793 f^4$\\
& $+ 48504 f^5 + 1416 f^6)$\\
\hline
\multirow{3}{*}{$d=3$} & $1/8(5778405000 + 3896455500 f$\\
& $+ 1138358475 f^2 + 184547230 f^3 + 17770209 f^4$\\
& $+ 988344 f^5 + 26584 f^6)$\\
\hline
\multirow{3}{*}{$d=4$} & $-1/12(11167090650 + 7366517730 f$\\
& $ + 2093789088 f^2 + 329260820 f^3 + 30595131 f^4$\\
& $+ 1626864 f^5 + 40440 f^6)$\\
\hline
\multirow{3}{*}{$d=5$} & $1/4(3139202700 + 2024766540 f$\\
& $+ 560010909 f^2 + 85380100 f^3 + 7645413 f^4$\\
& $+ 387408 f^5 + 8920 f^6)$\\
\hline
\multirow{3}{*}{$d=6$} & $-1/6(2602698750 + 1640270250 f$\\
& $+ 441280575 f^2 + 65148225 f^3 + 5609553 f^4 $\\
& $+ 270036 f^5 + 5768 f^6)$\\
\hline
\multirow{3}{*}{$d=7$} & $5/2(60727275 + 37359045 f$\\
& $+ 9765588 f^2 + 1393295 f^3 + 115023 f^4$\\
& $+ 5244 f^5 + 104 f^6)$\\
\hline
\multirow{3}{*}{$d=8$} & $-1/8(244861875 + 146850750 f$\\
& $ + 37228425 f^2 + 5118560 f^3 + 403710 f^4$\\
& $+ 17376 f^5 + 320 f^6)$\\
\hline
\multirow{3}{*}{$d=9$} & $1/24(65185425 + 38043810 f$\\
& $+ 9329139 f^2 + 1231680 f^3 + 92442 f^4$\\
& $+ 3744 f^5 + 64 f^6)$\\
\hline
\end{tabular}
\end{center}

\begin{center}
\begin{tabular}{| c || c |}
\hline
$\widetilde{(5,3)}$ & $GW_{(1,2)}(d,\vec{k}=\{2\})$ \\
\hline
\multirow{2}{*}{$d=0$} & $1/6(11370 + 4194 f + 596 f^2$ \\
& $+ 39 f^3 + f^4)$\\
\hline
\multirow{2}{*}{$d=1$} & $1/3(-48735 - 17769 f - 2545 f^2$\\
& $ - 174 f^3 - 5 f^4)$\\
\hline
\multirow{2}{*}{$d=2$} & $5/6(68250 + 24750 f + 3559 f^2$\\
& $+ 249 f^3 + 8 f^4)$\\
\hline
\multirow{2}{*}{$d=3$} & $1/3(-308400 - 111870 f - 16231 f^2$\\
& $- 1140 f^3 - 41 f^4)$\\
\hline
\multirow{2}{*}{$d=4$} & $5/24(484200 + 176190 f + 26099 f^2$\\
& $ + 1806 f^3 + 73 f^4)$\\
\hline
\multirow{2}{*}{$d=5$} & $-7/12(87420 + 31868 f + 4905 f^2$\\
& $ + 328 f^3 + 15 f^4)$\\
\hline
\multirow{2}{*}{$d=6$} & $7/24(35640 + 12934 f + 2117 f^2$ \\
& $ + 134 f^3 + 7 f^4)$\\
\hline
\end{tabular}
\end{center}

\begin{center}
\begin{tabular}{| c || c |}
\hline
$\widetilde{(5,3)}$ & $GW_{(1,3)}(d,\vec{k}=\{3\})$ \\
\hline
\multirow{3}{*}{$d=0$} & $-1/24(65185425 + 38043810 f$\\
& $+ 9329139 f^2 + 1231680 f^3 + 92442 f^4$\\
& $+ 3744 f^5 + 64 f^6)$\\
\hline
\multirow{3}{*}{$d=1$} & $1/8(244861875 + 146850750 f$\\
& $ + 37228425 f^2 + 5118560 f^3 + 403710 f^4$\\
& $+ 17376 f^5 + 320 f^6)$\\
\hline
\multirow{3}{*}{$d=2$} & $-5/2(60727275 + 37359045 f$\\
& $+ 9765588 f^2 + 1393295 f^3 + 115023 f^4$\\
& $+ 5244 f^5 + 104 f^6)$\\
\hline
\multirow{3}{*}{$d=3$} & $1/6(2602698750 + 1640270250 f$\\
& $+ 441280575 f^2 + 65148225 f^3 + 5609553 f^4 $\\
& $+ 270036 f^5 + 5768 f^6)$\\
\hline
\multirow{3}{*}{$d=4$} & $-1/4(3139202700 + 2024766540 f$\\
& $+ 560010909 f^2 + 85380100 f^3 + 7645413 f^4$\\
& $+ 387408 f^5 + 8920 f^6)$\\
\hline
\multirow{3}{*}{$d=5$} & $1/12(11167090650 + 7366517730 f$\\
& $ + 2093789088 f^2 + 329260820 f^3 + 30595131 f^4$\\
& $+ 1626864 f^5 + 40440 f^6)$\\
\hline
\multirow{3}{*}{$d=6$} & $-1/8(5778405000 + 3896455500 f$\\
& $+ 1138358475 f^2 + 184547230 f^3 + 17770209 f^4$\\
& $+ 988344 f^5 + 26584 f^6)$\\
\hline
\multirow{3}{*}{$d=7$} & $35/24 (242441100 + 167036040 f$\\
& $ + 50197806 f^2 + 8389630 f^3 + 836793 f^4$\\
& $+ 48504 f^5 + 1416 f^6)$\\
\hline
\multirow{3}{*}{$d=8$} & $-1/24(2375206875 + 1671279750 f$\\
& $+ 517317525 f^2 + 89166410 f^3 + 9219015 f^4$\\
& $ + 554568 f^5 + 17640 f^6)$\\
\hline
\multirow{3}{*}{$d=9$} & $1/24(289459575 + 207901890 f$\\
& $+ 66413862 f^2 + 11810810 f^3 + 1268583 f^4$ \\
& $+ 78792 f^5 + 2744 f^6)$\\
\hline
\end{tabular}
\end{center}

\begin{center}
\begin{tabular}{| c || c |}
\hline
$(5,3)$ & $GW_{(1,3)}(d,\vec{k}=\{2,1\})$ \\
\hline
\multirow{3}{*}{$d=0$} & $7/6(3292085250 + 3162917460 f$ \\
& $ + 1348506335 f^2 + 331925492 f^3 + 51255470 f^4$\\
& $+ 5040722 f^5 + 298753 f^6 + 8918 f^7)$\\
\hline
\multirow{4}{*}{$d=1$} & $1/6(-258325726500 - 241722288900 f$ \\ 
& $ - 100151504760 f^2 - 23899089127 f^3$\\
& $ - 3569763510 f^4 - 338147307 f^5$\\
& $ - 19200454 f^6 - 535080 f^7)$\\
\hline
\multirow{4}{*}{$d=2$} & $1/6(1306056365550 + 1190910846420 f$ \\
& $+ 479734400465 f^2 + 111038607209 f^3 $\\
& $ + 16046224025 f^4 + 1464335639 f^5$\\
& $+ 79586780 f^6 + 2076620 f^7)$\\
\hline
\multirow{4}{*}{$d=3$} & $1/6(-3940258725000 - 3502283143500 f$\\ 
& $- 1372081445850 f^2 - 308116450485 f^3$\\
& $- 43078759980 f^4 - 3786995917 f^5$\\
& $- 196875042 f^6 - 4821544 f^7)$\\
\hline
\multirow{4}{*}{$d=4$} & $1/6(7904012805000 + 6849173398500 f$ \\
& $+ 2609873144250 f^2 + 568634979615 f^3$\\
& $+ 76905742095 f^4 + 6510571023 f^5$ \\
& $+ 323569883 f^6 + 7453290 f^7)$\\
\hline
\multirow{4}{*}{$d=5$} & $-1/3(5555508832800 + 4693005675210 f$ \\
& $ + 1739189440795 f^2 + 367589667756 f^3$\\
& $+ 48073694455 f^4 + 3917245092 f^5$\\
& $+ 186024202 f^6 + 4037280 f^7)$\\
\hline
\multirow{4}{*}{$d=6$} & $2/3(2807162599500 + 2311119073200 f$\\ 
& $+ 832735208740 f^2 + 170668358436 f^3$\\
& $+ 21571039255 f^4 + 1690718132 f^5$\\
& $ + 76681787 f^6 + 1570244 f^7)$\\
\hline
\multirow{4}{*}{$d=7$} & $-1/3(4111848522000 + 3297814539930 f$\\
& $+ 1154732728161 f^2 + 229343805312 f^3$\\
& $+ 27992773785 f^4 + 2108736216 f^5$\\
& $+ 91296842 f^6 + 1765920 f^7)$\\
\hline
\multirow{4}{*}{$d=8$} & $1/6(4334617631250 + 3384499180500 f$\\
& $+ 1150816033275 f^2 + 221310397965 f^3$\\
& $+ 26060445345 f^4 + 1885046987 f^5$\\
& $+ 77861238 f^6 + 1423760 f^7)$\\
\hline
\multirow{4}{*}{$d=9$} & $-1/3(802352508750 + 609389602500 f$\\
& $+ 201025282525 f^2 + 37391295045 f^3$\\
& $+ 4242958105 f^4 + 294376219 f^5$\\
& $+ 11592672 f^6 + 200512 f^7)$\\
\hline
\multirow{4}{*}{$d=10$} & $1/2(132128464950 + 97511715300 f$\\
& $+ 31170910973 f^2 + 5600609493 f^3$\\
& $+ 611608465 f^4 + 40651643 f^5$\\
& $ + 1525160 f^6 + 24960 f^7)$\\
\hline
\multirow{3}{*}{$d=11$} & $-2/3(14677377750 + 10512388200 f$\\
& $+ 3251898175 f^2 + 563573721 f^3 + 59140915 f^4$\\
& $+ 3760951 f^5 + 134316 f^6 + 2080 f^7)$\\
\hline
\multirow{3}{*}{$d=12$} & $1/3(1973249100 + 1369667070 f $\\
& $+ 409378029 f^2 + 68322633 f^3 + 6878835 f^4$\\
& $+ 417961 f^5 + 14196 f^6 + 208 f^7)$\\
\hline
\end{tabular}
\end{center}

\begin{center}
\begin{tabular}{| c || c |}
\hline
$\widetilde{(5,3)}$ & $GW_{(1,3)}(d,\vec{k}=\{2,1\})$ \\
\hline
\multirow{3}{*}{$d=0$} & $-1/3(1973249100 + 1369667070 f $\\
& $+ 409378029 f^2 + 68322633 f^3 + 6878835 f^4$\\
& $+ 417961 f^5 + 14196 f^6 + 208 f^7)$\\
\hline
\multirow{3}{*}{$d=1$} & $2/3(14677377750 + 10512388200 f$\\
& $+ 3251898175 f^2 + 563573721 f^3 + 59140915 f^4$\\
& $+ 3760951 f^5 + 134316 f^6 + 2080 f^7)$\\
\hline
\multirow{4}{*}{$d=2$} & $-1/2(132128464950 + 97511715300 f$\\
& $+ 31170910973 f^2 + 5600609493 f^3$\\
& $+ 611608465 f^4 + 40651643 f^5$\\
& $ + 1525160 f^6 + 24960 f^7)$\\
\hline
\multirow{4}{*}{$d=3$} & $1/3(802352508750 + 609389602500 f$\\
& $+ 201025282525 f^2 + 37391295045 f^3$\\
& $+ 4242958105 f^4 + 294376219 f^5$\\
& $+ 11592672 f^6 + 200512 f^7)$\\
\hline
\multirow{4}{*}{$d=4$} & $-1/6(4334617631250 + 3384499180500 f$\\
& $+ 1150816033275 f^2 + 221310397965 f^3$\\
& $+ 26060445345 f^4 + 1885046987 f^5$\\
& $+ 77861238 f^6 + 1423760 f^7)$\\
\hline
\multirow{4}{*}{$d=5$} & $1/3(4111848522000 + 3297814539930 f$\\
& $+ 1154732728161 f^2 + 229343805312 f^3$\\
& $+ 27992773785 f^4 + 2108736216 f^5$\\
& $+ 91296842 f^6 + 1765920 f^7)$\\
\hline
\multirow{4}{*}{$d=6$} & $-2/3(2807162599500 + 2311119073200 f$\\ 
& $+ 832735208740 f^2 + 170668358436 f^3$\\
& $+ 21571039255 f^4 + 1690718132 f^5$\\
& $ + 76681787 f^6 + 1570244 f^7)$\\
\hline
\multirow{4}{*}{$d=7$} & $1/3(5555508832800 + 4693005675210 f$ \\
& $ + 1739189440795 f^2 + 367589667756 f^3$\\
& $+ 48073694455 f^4 + 3917245092 f^5$\\
& $+ 186024202 f^6 + 4037280 f^7)$\\
\hline
\multirow{4}{*}{$d=8$} & $-1/6(7904012805000 + 6849173398500 f$ \\
& $+ 2609873144250 f^2 + 568634979615 f^3$\\
& $+ 76905742095 f^4 + 6510571023 f^5$ \\
& $+ 323569883 f^6 + 7453290 f^7)$\\
\hline
\multirow{4}{*}{$d=9$} & $1/6(3940258725000 + 3502283143500 f$\\ 
& $+ 1372081445850 f^2 + 308116450485 f^3$\\
& $+ 43078759980 f^4 + 3786995917 f^5$\\
& $+ 196875042 f^6 + 4821544 f^7)$\\
\hline
\multirow{4}{*}{$d=10$} & $-1/6(1306056365550 + 1190910846420 f$ \\
& $+ 479734400465 f^2 + 111038607209 f^3 $\\
& $ + 16046224025 f^4 + 1464335639 f^5$\\
& $+ 79586780 f^6 + 2076620 f^7)$\\
\hline
\multirow{4}{*}{$d=11$} & $1/6(258325726500 + 241722288900 f$ \\ 
& $ + 100151504760 f^2 + 23899089127 f^3$\\
& $ + 3569763510 f^4 + 338147307 f^5$\\
& $ + 19200454 f^6 + 535080 f^7)$\\
\hline
\multirow{3}{*}{$d=12$} & $-7/6(3292085250 + 3162917460 f$ \\
& $ + 1348506335 f^2 + 331925492 f^3 + 51255470 f^4$\\
& $+ 5040722 f^5 + 298753 f^6 + 8918 f^7)$\\
\hline
\end{tabular}
\end{center}


\begin{center}
\begin{tabular}{| c || c |}
\hline
$(5,3)$ & $GW_{(2,2)}(d,\vec{k}=\{2\})$ \\
\hline
\multirow{2}{*}{$d=0$} & $7/1440(10757880 + 5188718 f + 1144821 f^2$ \\
& $+ 129340 f^3 + 10585 f^4 + 402 f^5 + 14 f^6)$\\
\hline
\multirow{2}{*}{$d=1$} & $-7/720(26668140 + 12876556 f + 2750805 f^2$\\
& $+ 318680 f^3 + 24525 f^4 + 984 f^5 + 30 f^6)$\\
\hline
\multirow{2}{*}{$d=2$} & $1/288(146770920 + 71037414 f + 14988891 f^2$\\
& $ + 1761900 f^3 + 130495 f^4 + 5418 f^5 + 146 f^6)$\\
\hline
\multirow{2}{*}{$d=3$} & $1/180(-91423200 - 44592510 f - 9427263 f^2$\\
& $- 1118700 f^3 - 81155 f^4 - 3420 f^5 - 82 f^6)$\\
\hline
\multirow{2}{*}{$d=4$} & $1/72(19472730 + 9656076 f + 2067216 f^2$\\
& $+ 247500 f^3 + 17795 f^4 + 747 f^5 + 16 f^6)$\\
\hline
\multirow{2}{*}{$d=5$} & $1/180(-13153995 - 6708573 f - 1467285 f^2$\\
& $- 177690 f^3 - 12725 f^4 - 522 f^5 - 10 f^6)$\\
\hline
\multirow{2}{*}{$d=6$} & $1/360(2839290 + 1512708 f + 341283 f^2$\\
& $+ 41940 f^3 + 2980 f^4 + 117 f^5 + 2 f^6)$\\
\hline
\end{tabular}
\end{center}

\begin{center}
\begin{tabular}{| c || c |}
\hline
$(5,3)$ & $GW_{(2,3)}(d,\vec{k}=\{3\})$ \\
\hline
\multirow{3}{*}{$d=0$} & $\hbox{\fontsize{7}{15}\selectfont\(1/1920(-313660234575 - 279376532190 f\)}$\\
& $\hbox{\fontsize{7}{15}\selectfont\( - 113040123102 f^2 - 26943588238 f^3- 4185580751 f^4\)}$ \\
& $\hbox{\fontsize{7}{15}\selectfont\( - 437194160 f^5- 31064176 f^6 - 1365728 f^7 - 35672 f^8)\)}$\\
\hline
\multirow{3}{*}{$d=1$} & $\hbox{\fontsize{7}{15}\selectfont\( 1/1920(2681338966875 + 2339120860650 f\)}$\\
& $\hbox{\fontsize{7}{15}\selectfont\(  + 922575016395 f^2 + 214387842382 f^3 + 32399106105 f^4\)}$\\
& $\hbox{\fontsize{7}{15}\selectfont\(+ 3293913200 f^5 + 225391680 f^6 + 9612512 f^7 + 229320 f^8)\)}$\\
\hline
\multirow{3}{*}{$d=2$} & $\hbox{\fontsize{7}{15}\selectfont\(-7/384(283634613900 + 242690987160 f\)}$\\
& $\hbox{\fontsize{7}{15}\selectfont\(+ 93541999554 f^2 + 21228329714 f^3 + 3126888057 f^4 \)}$\\
& $\hbox{\fontsize{7}{15}\selectfont\( + 309302320 f^5 + 20430096 f^6 + 840736 f^7 + 18408 f^8)\)}$\\
\hline
\multirow{3}{*}{$d=3$} & $\hbox{\fontsize{7}{15}\selectfont\(1/640(6977259864000 + 5862908261700 f\)}$\\
& $\hbox{\fontsize{7}{15}\selectfont\(+ 2212674250545 f^2 + 491050156226 f^3 + 70575897711 f^4\)}$\\
& $\hbox{\fontsize{7}{15}\selectfont\( + 6790495600 f^5 + 433333120 f^6 + 17131296 f^7 + 345592 f^8)\)}$\\
\hline
\multirow{3}{*}{$d=4$} & $\hbox{\fontsize{7}{15}\selectfont\(-1/960(13874376720150 + 11461626400230 f\)}$\\
& $\hbox{\fontsize{7}{15}\selectfont\(+ 4241680714008 f^2 + 921402229396 f^3 + 129271155135 f^4\)}$\\
& $\hbox{\fontsize{7}{15}\selectfont\(+ 12092640320 f^5 + 745295184 f^6 + 28198976 f^7 + 525720 f^8)\)}$\\
\hline
\multirow{3}{*}{$d=5$} & $\hbox{\fontsize{7}{15}\selectfont\(1/320(4005018524700 + 3255636685140 f\)}$\\
& $\hbox{\fontsize{7}{15}\selectfont\(+ 1182625926459 f^2 + 251566127852 f^3 + 34447385255 f^4\)}$\\
& $\hbox{\fontsize{7}{15}\selectfont\( + 3130161760 f^5 + 186073952 f^6 + 6715072 f^7 + 115960 f^8)\)}$\\
\hline
\multirow{3}{*}{$d=6$} & $\hbox{\fontsize{7}{15}\selectfont\(-1/480(3406361559750 + 2726585335350 f\)}$\\
& $\hbox{\fontsize{7}{15}\selectfont\(+ 972679182885 f^2 + 202590159819 f^3 + 27054228137 f^4\)}$\\
& $\hbox{\fontsize{7}{15}\selectfont\( + 2384464200 f^5 + 136419520 f^6 + 4680624 f^7 + 74984 f^8)\)}$\\
\hline
\multirow{3}{*}{$d=7$} & $\hbox{\fontsize{7}{15}\selectfont\(1/32(81539428725 + 64294728255 f\)}$\\
& $\hbox{\fontsize{7}{15}\selectfont\(+22524635112 f^2 + 4590237529 f^3 + 596935513 f^4\)}$\\
& $\hbox{\fontsize{7}{15}\selectfont\(+50917160 f^5 + 2795408 f^6 + 90896 f^7 + 1352 f^8)\)}$\\
\hline
\multirow{3}{*}{$d=8$} & $\hbox{\fontsize{7}{15}\selectfont\(-1/640(337678963875 + 262331586450 f\)}$\\
& $\hbox{\fontsize{7}{15}\selectfont\(+90204481335 f^2 + 17961279064 f^3 + 2269578340 f^4\)}$\\
& $\hbox{\fontsize{7}{15}\selectfont\(+186792800 f^5 + 9805840 f^6 + 301184 f^7 + 4160 f^8)\)}$\\
\hline
\multirow{3}{*}{$d=9$} & $\hbox{\fontsize{7}{15}\selectfont\(1/1920(92512861425 + 70800532110 f\)}$\\
& $\hbox{\fontsize{7}{15}\selectfont\(+23870277549 f^2 + 4634733576 f^3 + 567379036 f^4\)}$\\
& $\hbox{\fontsize{7}{15}\selectfont\(+44891040 f^5 + 2244272 f^6 + 64896 f^7 + 832 f^8)\)}$\\
\hline
\end{tabular}
\end{center}

\begin{center}
\begin{tabular}{| c || c |}
\hline
$\widetilde{(5,3)}$ & $GW_{(2,2)}(d,\vec{k}=\{2\})$ \\
\hline
\multirow{2}{*}{$d=0$} & $1/360(2839290 + 1512708 f + 341283 f^2$ \\
& $+ 41940 f^3 + 2980 f^4 + 117 f^5 + 2 f^6)$\\
\hline
\multirow{2}{*}{$d=1$} & $1/180(-13153995 - 6708573 f - 1467285 f^2$\\
& $- 177690 f^3 - 12725 f^4 - 522 f^5 - 10 f^6)$\\
\hline
\multirow{2}{*}{$d=2$} & $1/72(19472730 + 9656076 f + 2067216 f^2$\\
& $+ 247500 f^3 + 17795 f^4 + 747 f^5 + 16 f^6)$\\
\hline
\multirow{2}{*}{$d=3$} & $1/180(-91423200 - 44592510 f - 9427263 f^2$\\
& $- 1118700 f^3 - 81155 f^4 - 3420 f^5 - 82 f^6)$\\
\hline
\multirow{2}{*}{$d=4$} & $1/288(146770920 + 71037414 f + 14988891 f^2$\\
& $+ 1761900 f^3 + 130495 f^4 + 5418 f^5 + 146 f^6)$\\
\hline
\multirow{2}{*}{$d=5$} & $-7/720(26668140 + 12876556 f + 2750805 f^2$\\
& $+ 318680 f^3 + 24525 f^4 + 984 f^5 + 30 f^6)$\\
\hline
\multirow{2}{*}{$d=6$} & $7/1440(10757880 + 5188718 f + 1144821 f^2$\\
& $+ 129340 f^3 + 10585 f^4 + 402 f^5 + 14 f^6)$\\
\hline
\end{tabular}
\end{center}

\vskip 0.87cm

\begin{center}
\begin{tabular}{| c || c |}
\hline
$\widetilde{(5,3)}$ & $GW_{(2,3)}(d,\vec{k}=\{3\})$ \\
\hline
\multirow{3}{*}{$d=0$} & $\hbox{\fontsize{7}{15}\selectfont\(-1/1920(92512861425 + 70800532110 f\)}$\\
& $\hbox{\fontsize{7}{15}\selectfont\(+23870277549 f^2 + 4634733576 f^3 + 567379036 f^4\)}$\\
& $\hbox{\fontsize{7}{15}\selectfont\(+44891040 f^5 + 2244272 f^6 + 64896 f^7 + 832 f^8)\)}$\\
\hline
\multirow{3}{*}{$d=1$} & $\hbox{\fontsize{7}{15}\selectfont\(1/640(337678963875 + 262331586450 f\)}$\\
& $\hbox{\fontsize{7}{15}\selectfont\(+90204481335 f^2 + 17961279064 f^3 + 2269578340 f^4\)}$\\
& $\hbox{\fontsize{7}{15}\selectfont\(+186792800 f^5 + 9805840 f^6 + 301184 f^7 + 4160 f^8)\)}$\\
\hline
\multirow{3}{*}{$d=2$} & $\hbox{\fontsize{7}{15}\selectfont\(-1/32(81539428725 + 64294728255 f\)}$\\
& $\hbox{\fontsize{7}{15}\selectfont\(+22524635112 f^2 + 4590237529 f^3 + 596935513 f^4\)}$\\
& $\hbox{\fontsize{7}{15}\selectfont\(+50917160 f^5 + 2795408 f^6 + 90896 f^7 + 1352 f^8)\)}$\\
\hline
\multirow{3}{*}{$d=3$} & $\hbox{\fontsize{7}{15}\selectfont\(1/480(3406361559750 + 2726585335350 f\)}$\\
& $\hbox{\fontsize{7}{15}\selectfont\(+ 972679182885 f^2 + 202590159819 f^3 + 27054228137 f^4\)}$\\
& $\hbox{\fontsize{7}{15}\selectfont\( + 2384464200 f^5 + 136419520 f^6 + 4680624 f^7 + 74984 f^8)\)}$\\
\hline
\multirow{3}{*}{$d=4$} & $\hbox{\fontsize{7}{15}\selectfont\(-1/320(4005018524700 + 3255636685140 f\)}$\\
& $\hbox{\fontsize{7}{15}\selectfont\(+ 1182625926459 f^2 + 251566127852 f^3 + 34447385255 f^4\)}$\\
& $\hbox{\fontsize{7}{15}\selectfont\( + 3130161760 f^5 + 186073952 f^6 + 6715072 f^7 + 115960 f^8)\)}$\\
\hline
\multirow{3}{*}{$d=5$} & $\hbox{\fontsize{7}{15}\selectfont\(1/960(13874376720150 + 11461626400230 f\)}$\\
& $\hbox{\fontsize{7}{15}\selectfont\(+ 4241680714008 f^2 + 921402229396 f^3 + 129271155135 f^4\)}$\\
& $\hbox{\fontsize{7}{15}\selectfont\(+ 12092640320 f^5 + 745295184 f^6 + 28198976 f^7 + 525720 f^8)\)}$\\
\hline
\multirow{3}{*}{$d=6$} & $\hbox{\fontsize{7}{15}\selectfont\(-1/640(6977259864000 + 5862908261700 f\)}$\\
& $\hbox{\fontsize{7}{15}\selectfont\(+ 2212674250545 f^2 + 491050156226 f^3 + 70575897711 f^4\)}$\\
& $\hbox{\fontsize{7}{15}\selectfont\( + 6790495600 f^5 + 433333120 f^6 + 17131296 f^7 + 345592 f^8)\)}$\\
\hline
\multirow{3}{*}{$d=7$} & $\hbox{\fontsize{7}{15}\selectfont\(7/384(283634613900 + 242690987160 f\)}$\\
& $\hbox{\fontsize{7}{15}\selectfont\(+ 93541999554 f^2 + 21228329714 f^3 + 3126888057 f^4 \)}$\\
& $\hbox{\fontsize{7}{15}\selectfont\( + 309302320 f^5 + 20430096 f^6 + 840736 f^7 + 18408 f^8)\)}$\\
\hline
\multirow{3}{*}{$d=8$} & $\hbox{\fontsize{7}{15}\selectfont\(-1/1920(2681338966875 + 2339120860650 f\)}$\\
& $\hbox{\fontsize{7}{15}\selectfont\(  + 922575016395 f^2 + 214387842382 f^3 + 32399106105 f^4\)}$\\
& $\hbox{\fontsize{7}{15}\selectfont\(+ 3293913200 f^5 + 225391680 f^6 + 9612512 f^7 + 229320 f^8)\)}$\\
\hline
\multirow{3}{*}{$d=9$} & $\hbox{\fontsize{7}{15}\selectfont\(1/1920(313660234575 + 279376532190 f\)}$\\
& $\hbox{\fontsize{7}{15}\selectfont\( + 113040123102 f^2 + 26943588238 f^3 + 4185580751 f^4\)}$ \\
& $\hbox{\fontsize{7}{15}\selectfont\( + 437194160 f^5 + 31064176 f^6 + 1365728 f^7 + 35672 f^8)\)}$\\
\hline
\end{tabular}
\end{center}

\end{document}